\documentclass[nocompress]{spie}  

\usepackage{amsmath,amsfonts,amssymb}
\def\arcmin{\hbox{$^\prime$}}
\def\arcsec{\hbox{$^{\prime\prime}$}}
\def\degrees{\ifmmode^\circ\else$^\circ$\fi}
\usepackage{graphicx}
\usepackage{hyperref}
\hypersetup
{
colorlinks=true
allcolors={blue}
}

\title{Current Status of the Facility Instrumentation Suite at The Large Binocular Telescope Observatory}

\author[a]{Barry Rothberg}
\author[a]{Olga Kuhn}
\author[a]{Michelle L. Edwards}
\author[a]{John M. Hill}
\author[a]{David Thompson}
\author[a]{Christian Veillet}
\author[a]{R. Mark Wagner}

\affil[a]{Large Binocular Telescope Observatory, 933 North Cherry Avenue, Tucson, AZ 85721,USA}

\authorinfo{Further author information: (Send correspondence to Barry Rothberg)\\E-mail: brothberg@lbto.org, Telephone: +1 520 626-8672, Fax: +1 520 626-9333}

\begin{document} 
\maketitle

\begin{abstract}
The current status of the facility instrumentation for the Large Binocular Telescope 
(LBT) is reviewed. The LBT encompasses two 8.4 meter primary mirrors on a single mount 
yielding an effective collecting area of 11.8 meters or 23 meters when interferometrically 
combined. The three facility instruments at LBT include: 1) the Large Binocular Cameras 
(LBCs), each with a 23\arcmin $\times$ 25\arcmin\ field of view (FOV). The blue optimized 
and red optimized optical wavelength LBCs are mounted at the prime focus of the SX (left) 
and DX (right) primary mirrors, respectively. Combined, the filter suite of the two LBCs 
cover 0.3-1.1 $\mu$m, including the addition of new medium-band filters centered on TiO 
(0.78 $\mu$m) and CN (0.82 $\mu$m); 2) the Multi-Object Double Spectrograph (MODS), two 
identical optical spectrographs each mounted at the straight through f/15 Gregorian focus 
of the primary mirrors. The capabilities of MODS-1 and -2 include imaging with Sloan 
filters ({\it u, g, r, i, and z}) and medium resolution ({\it R} $\sim$ 2000) spectroscopy, 
each with 24 interchangeable masks (multi-object or longslit) over a 6\arcmin $\times$
6\arcmin\ FOV. Each MODS 
is capable of blue (0.32-0.6 $\mu$m) and red (0.5-1.05 $\mu$m) wavelength only 
spectroscopy coverage or both can employ a dichroic for 0.32-1.05 $\mu$m wavelength 
coverage (with reduced coverage from 0.56-0.57 $\mu$m); and 3) the two LBT Utility 
Camera in the Infrared instruments (LUCIs), are each mounted at a bent-front 
Gregorian f/15 focus of a primary mirror. LUCI-1 \& 2 are designed for seeing-limited 
(4\arcmin $\times$ 4\arcmin\ FOV) and active optics using thin-shell adaptive secondary 
mirrors (0.5\arcmin $\times$ 0.5\arcmin\ FOV) imaging and spectroscopy over the wavelength 
range of 0.95-2.5 $\mu$m and spectroscopic resolutions of 400 $\le$ {\it R} $\le$ 11000
(depending on the combination of grating, slits, and cameras used).
The spectroscopic capabilities also include 32 interchangeable multi-object or longslit 
masks which are cryogenically cooled.  Currently all facility instruments are in-place at 
the LBT and, for the first time, have been on-sky for science observations. In Summer 2015 
LUCI-1 was refurbished to replace the infrared detector; to install a high-resolution 
camera to take advantage of the active optics SX secondary; and to install a 
grating designed primarily for use with high resolution active optics. Thus, like MODS-1 
\& -2, both LUCIs now have specifications nearly identical 
to each other. The software interface for both LUCIs have also been replaced, allowing 
both instruments to be run together from a single interface.  With the installation of 
all facility instruments finally complete we also report on the first science use of 
``mixed-mode'' operations, defined as the combination of different paired instruments 
with each mirror (i.e. LBC+MODS, LBC+LUCI, LUCI+MODS). Although both primary mirrors 
reside on a single fixed mount, they are capable of operating as independent entities 
within a defined ``co-pointing'' limit. This provides users with the additional capability 
to independently dither each mirror or center observations on two different sets of 
spatial coordinates within this limit.
\end{abstract}


\section{INTRODUCTION}
\label{sec:intro}  
The Large Binocular Telescope Observatory (LBTO) is situated near the city of Safford in
southeastern Arizona in the Pinale\~{n}o Mountains. It is part of the Mount Graham 
International Observatory (MGIO) located on Emerald Peak on the highest mountain, 
Mount Graham.  LBTO sits at an altitude of 3192 meters.  As the name implies, the Large 
Binocular Telescope (LBT) houses two primary mirrors, separated by 14.4 meters 
(center-to-center), mounted on a single altitude-azimuth mount.  Each mirror is 8.4 meters 
in diameter, with a combined collecting area equivalent to a single 11.8 meter mirror, 
or an interferometric baseline of 22.65 meters, edge-to-edge.  Although the LBT is 
comprised of two 8.4 meter mirrors, their fast {\it f/1.14} focal ratio allows for a
compact mount and co-rotating enclosure, see Hill et al. (2004)\cite{2004SPIE.5489..603H},
Ashby et al. (2006)\cite{2006SPIE.6274E..23A}, 
Hill et al. (2006)\cite{2006SPIE.6267E..0YH}
and Hill et al. (2010)\cite{2010SPIE.7733E..0CH} for more details.  The binocular design 
is combined with four Bent Gregorian focal stations (three with instrument rotator 
bearings) and one Direct Gregorian focal station 
for each side of the telescope.  The two mirrors can be used in binocular mode with
the same pairs of instruments as well as each independently with different 
instrumentation.  Switching between different optical instrumentation is done by moving 
various swing arms which hold the prime focus optical cameras, or secondary and tertiary 
mirrors.  The transition between prime focus and Gregorian instruments takes $\sim$ 20 
minutes, while transitions between different Gregorian instruments can take 
$\le$ 10 minutes.  This flexibility is advantageous for incorporating a variety of 
scientific programs during a single night as well as adapting quickly to changes in site 
and weather conditions.  For brevity, the left-side of the telescope is denoted as 
{\it SX} and the right-side is denoted as {\it DX}.\\
\indent The LBT is an international partnership which includes the University of Arizona 
(25$\%$) including access for Arizona State University and Northern Arizona University;
Germany (25$\%$) or LBTB (Beteiligungsgesellschaft) which includes participation of five 
German institutes (Landessternwarte K{\"{o}nigstuhl, Leibniz Institute for Astrophysics
Potsdam, Max-Planck-Institut f{\"{u}r Astronomie, Max-Planck-Institut f{\"{u}r 
Extraterrestrische Physik, and Max-Planck-Institut f{\"{u}r Radioastronomie); Italy 
(25$\%$) or Instituto Nazionale di Astrofisica which is responsible for offering 
access to the Italian community to LBTO; the Ohio State University (12.5$\%$); and
the Research Corporation for Science and Advancement (RC) which coordinates
the participation of four universities (Ohio State University, University of Notre Dame,
University of Minnesota, and University of Virginia).\\
\indent In 2014, the second Multi-Object Double Spectrograph (MODS-2) was installed
on the LBT.  This marked the completion of the installation of all facility 
instrumentation.  Beginning with the 2014B semester and up through the 2016A semester, 
all facility instruments (3 on SX, 3 on DX) were available for on-sky scientific use 
during partner and LBTO science time or underwent commissioning (or re-commissioning). 
For the 2016B semester, all six facility instruments will be available for on-sky 
scientific use by the LBTO and its partners.  In this conference proceeding, we present
a summary of the LBT scientific facility instruments that are now available for
partner science observations.  It is an update on the significant changes that have
occurred at LBTO since Wagner et al. (2014)\cite{2014SPIE.9147E..05W}.

\section{Types of Instrumentation}
\label{sec:types} 
There are three categories of LBT scientific instrumentation.  The first are {\it facility
instruments}, which are available for use by anyone within the partnerships. Facility 
instruments are supported and maintained by LBTO personnel.  Although during commissioning
phases, facility instruments are still supported by the instrument teams, who work in 
conjunction with LBTO staff.  The second are {\it Principal Investigator instruments} such 
as the Potsdam Echelle and Polarimetric Spectroscopic Instrument (PEPSI), which uses
both primary mirrors, see Strassmeier et al. (2008)\cite{2008SPIE.7014E..0NS} for more 
information, and has been 
used on-sky for scientific observations during the 2015B and 2016A semesters.  These are 
maintained and operated solely by the builders, but may be used by LBTO partners for 
science on a collaborative basis or through time exchanges at the discretion of the 
instrument principal investigator (PI). LBTO provides limited technical assistance to 
enable the instruments to interface with the telescope control systems and telescope 
infrastructure.  The third type are {\it Strategic instruments}, which are technically 
challenging and designed to push the limits of astronomical instrumentation.  They may 
be uniquely suited to the LBT and are designed to have a major impact on astronomy.  
Strategic instruments may be available to the LBT community on a collaborative basis or 
through time exchanges.  A current strategic instrument is the 
LBT Interferometer (LBTI), which uses both primary mirrors and comprises 
LMIRCam (3-5 $\mu$m) and the NOMIC (8-13$\mu$m) camera. They are currently operational 
for on-sky scientific observations, see Hinz et al. (2008)\cite{2008SPIE.7013E..28H},
Wilson et al. (2008)\cite{2008SPIE.7013E..3AW}, 
Skrutskie et al. 2010\cite{2010SPIE.7735E..3HS}, 
Leisenring et al. (2012)\cite{2012SPIE.8446E..4FL}, 
and Hoffmann et al. (2014)\cite{2014SPIE.9147E..1OH} for more information.  
Recent results include mapping the 
5 $\mu$m emission of the Loki Patera volcanoes on Jupiter's moon Io using the 
SX and DX mirrors in interferometric mode (Conrad et al. 2015\cite{2015AJ....149..175C})
or the LBTI Exozodi Exoplanet Common Hunt (LEECH) survey 
(Skemer et al. 2016\cite{2016ApJ...817..166S}).
Future strategic instruments include: LBT INterferometric Camera and the Near–IR/Visible 
Adaptive iNterferometer for Astronomy (LINC- NIRVANA), a multi-conjugate adaptive optics
(MCAO) near-IR imaging system that provides both ground-layer and high-layer corrections.
It is in the first stages of on-telescope testing 
(Gassler et al. (2004)\cite{2004SPIE.5382..742G},  
Herbst et al. (2014)\cite{2014SPIE.9147E..1MH}); and iLocater, a 
diffraction-limited Doppler spectrometer with high spectral resolution 
({\it R} $\sim$ 110,000) operating in the {\it Y}-band that will be used to characterize 
Earth-like exo-planets orbiting M-dwarf stars.  iLocater will use a fiber-fed AO-corrected 
beam to pass light to a compact spectrograph (Crepp at al. 2014\cite{2014AAS...22334820C},
Veillet et al. 2014\cite{2014SPIE.9149E..16V}).  Further discussion of PI and 
Strategic instruments are beyond the scope of this paper.  

\section{Facility Instruments}
\label{sec:facility} 
The Large Binocular Cameras (LBCs) are a pair of blue- and red-optimized prime focus 
imagers with a field of view just shy the size of the Moon projected on the sky.  The LBC 
filter suite covers {\it U}-band (0.33 $\mu$m) in the blue through {\it Y}-band (1.1 $\mu$m)
in the red.  The Multi-Object Double Spectrographs 1 and 2 (MODS-1 and MODS-2) are a pair of 
optical spectrographs (longslit and custom located at the 
direct Gregorian foci at each primary mirror.  The two LBT NIR Spectroscopic Utility with 
Camera Instruments (LUCIs) are capable of imaging and 
spectroscopy (longslit and custom designed multi-object slit masks) each located at one 
of the Bent {\it f/15} Gregorian ports.  As of the 2016A semester all facility instruments have 
been used for scientific observations on-sky. Table 1 presents a brief overview of the 
scientific capabilities of the facility instruments.  
Specific details will be discussed or cited in subsequent sections.   

\begin{table}[ht]
\caption{Overview of Facility Instruments \& Capabilities} 
\label{tab:FacCap}
\begin{center}       
\begin{tabular}{|l|l|l|l|l|l|l|} 
\hline
\rule[-1ex]{0pt}{3.5ex}   {\bf Instrument} & {\bf Focal} & {\bf MODES}  &{\bf $\lambda$ Range} &{\bf FOV}  &{\bf Plate Scale}  &{\bf Resolution}\\
\rule[-1ex]{0pt}{3.5ex}    &  {\bf Station}&      &($\mu$m) &(\arcmin)               & (\arcsec/pixel)     & \\
\hline
\rule[-1ex]{0pt}{3.5ex}   LBC Blue  & SX Prime      & Imaging  & 0.33-0.67  & 23\arcmin $\times$ 25\arcmin & 0.2255    & ...     \\
\hline
\rule[-1ex]{0pt}{3.5ex}   LBC Red   & DX Prime      & Imaging  & 0.55-1.11  & 23\arcmin $\times$ 25\arcmin & 0.2255    & ...     \\
\hline
\rule[-1ex]{0pt}{3.5ex}   MODS-1/-2  & {\it f/15} Direct & Imaging   & 0.31-1.1 & 6\arcmin $\times$ 6\arcmin & 0.120 {\footnotesize(blue)} & {\footnotesize2300 (blue)} \\
\rule[-1ex]{0pt}{3.5ex}              & Gregorian   & Spectroscopy &          &                   & 0.123 {\footnotesize(red)}  & {\footnotesize1850 (red)}\\
\rule[-1ex]{0pt}{3.5ex}              &             &              &          &                   &              &{\footnotesize(100-500 prism)}\\
\hline
\rule[-1ex]{0pt}{3.5ex}   LUCI-1/-2  &{\it f/15} Bent   &Imaging       & 0.95-2.5  & 4\arcmin $\times$ 4\arcmin & 0.25 {\footnotesize(N1.8)} & \footnotesize1500-5500 {(N1.8)}    \\
\rule[-1ex]{0pt}{3.5ex}   {\footnotesize(Seeing Limited)} &Greogorian  &Spectroscopy  &           &                 & 0.12 {\footnotesize(N3.75)}& {\footnotesize3000-11,000 (N3.75)} \\
\hline
\end{tabular}
\end{center}
\end{table}

\subsection{Large Binocular Cameras (LBCs)}
\label{subsec:LBCs}
The LBCs are comprised of two wide-field imagers, one blue optimized at the {\it f/1.14} 
prime focus on SX, and one red optimized at the {\it f/1.14} prime focus on DX.  They 
are each mounted on a spider swing-arm that can be deployed above their
respective primary mirror and moved into and out of the telescope beam as required.  
The LBCs were accepted as facility instruments in October 2011.  The two
instruments were an in-kind
contribution by INAF to the first generation of LBT instruments.  
Specific details regarding construction, commissioning, and upgrades 
can be found in Ragazzoni et al. (2006)\cite{2006SPIE.6267E..10R},
Speziali et al. (2008)\cite{2008SPIE.7014E..4TS}, and 
Giallongo et al. (2008)\cite{2008A&A...482..349G}. 
The LBCs are the first instruments to make full use of the binocular capabilities
of the LBT.  Binocular observing has been done with the LBCs since the installation
and commissioning of LBC Red was completed.\\
\indent Owing to the fast focal ratio of the primary mirrors and placement at prime focus, 
a set of refractor corrector lenses is required to deal with geometric distortions that 
would affect the large field of view (FOV). Each LBC uses a similar set of five 
corrective lens (a 6th lens is the cryostat window with almost no net power).  
This is based on the Wynne approach of positive-negative-positive lenses 
(Wynne 1972\cite{1972MNRAS.160P..13W}), but with the second and third elements each split 
into two lenses. A filter wheel sits between the 5$^{th}$ and 6$^{th}$ corrective lens
(the first lens is defined as closest to the primary mirror).  The lenses in LBC Blue are 
made of fused silica which 
permits better transmittance of light at shorter wavelengths (0.3-0.5 $\mu$m).  The 
lenses in LBC Red use borosilicate glass (BK7) and are optimized for longer wavelengths 
($\lambda$ $>$ 0.5 $\mu$m).  The corrected fields have a diameter 
of 110 mm and 108.2 mm for LBC Blue and LBC Red (equivalent to 27\arcmin\  in diameter), 
respectively.  The science detectors cover $\sim$ 75$\%$ of this area.\\
\indent The LBCs each contain six E2V CCD detectors, four of which are used for science. 
The four science CCDs are E2V 420-90s with 2048 $\times$ 4608 (13.5 $\mu$m square pixels) 
are arranged in a mosaic with three abutted next to each other.  A fourth CCD is rotated 
clockwise 90 degrees and centered along the top of the three science CCDs.  Each CCD 
covers 7\arcmin.8 $\times$ 17\arcmin.6 with a gap of 70 pixels (18\arcsec\ ) between each 
CCD.  This yields a 23\arcmin\ $\times$ 25\arcmin\ FOV. In order to obtain an 
uninterrupted image, dithering is required to to fill the gaps between CCDs 
(and recommended to correct for cosmic rays and bad pixel columns). The un-binned
readout time for the full array of science CCDs is 27 seconds.  The other two CCDs 
are used for guiding and tracking collimation and wavefront control.  They are 
E2V 420-90 custom made 512 $\times$ 2048 (13.5 $\mu$m square) pixel CCDs that do not 
have a shutter mechanism. They are placed on either of the science CCD chips.  One is 
within the focal plane and is used for guiding adjustments, the other is out of the focal 
plane and uses extra-focal pupil images to maintain collimation and focus.  Figure 1 shows
the layout of LBC Blue (LBC Red is the same layout with a slight difference in the 
corrected field size), computed distortion map, and an example of a deep UV image.

   \begin{figure} [ht]
   \begin{center}
   \begin{tabular}{c} 
   \includegraphics[height=6cm]{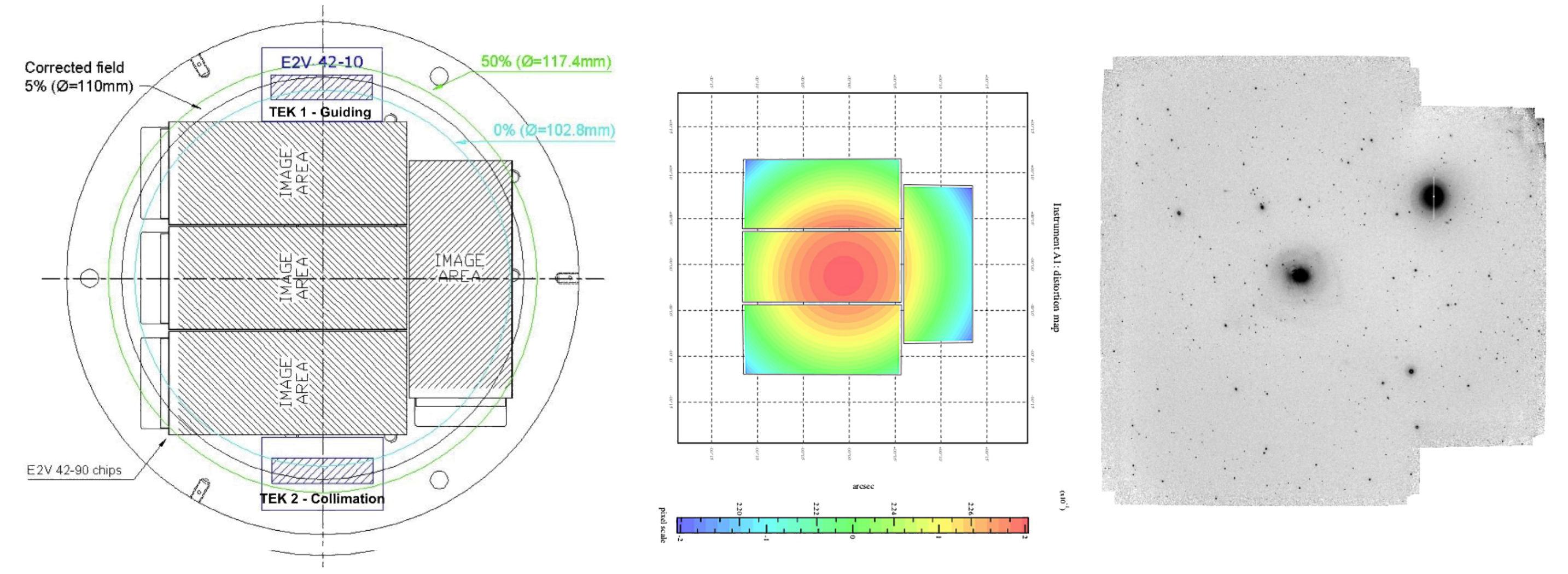}
   \end{tabular}
   \end{center}
   Figure 1:  Shown here are: a) the chip layout of LBC Blue (science and technical chips);
   b) Distortion map computed for LBC Blue; and c) 2160 sec {\it U}-spec image of the 
   galaxy merger.  The project goal
   is to map the extent of the Globular Cluster (GC) and Young Star Cluster (YSC) 
   population, which appear as partially or unresolved point sources across the field.  
   Here, the LBC FOV covers $\sim$ 160 $\times$ 150 metric kiloparsecs, which should cover the
   entire spatial extent of the GC and YSC populations 
   (Rochais et al.\cite{2016AAS...22724012R}).
    \end{figure} 

\indent Each of the LBCs houses two filter wheels, each wheel houses 5 slots, which allows
for up to 8 filters to be used for each instrument (one slot in each wheel 
must always be empty).  Although nominally of similar design, the LBCs use 
different filters of different widths.  LBC Blue filters are 159.8 mm in diameter 
(155 mm opening), while LBC Red filters are 189.6 mm in diameter (185 mm opening).
Currently, LBC Blue houses six filters for scientific use:  {\it U}-Bessel, 
{\it U}-Spectroscopic (broader transmission curve than {\it U}-Bessel), {\it B}-Bessel, 
{\it V}-Bessel, Sloan {\it g}, {\it r}. There are now ten filters available for 
scientific use with LBC Red:  {\it V}-Bessel, {\it R}-Bessel, {\it I}-Bessel, 
sloan {\it r}, {\it i}, and {\it z}, and {\it Y}-band filter; and 
three medium filters, F972N20, TiO 784 and CN 817.  The TiO 784 and CN 817 filters were 
purchased and tested in semester 2014B by Landessternwarte K{\"{o}nigstuhl (LBTB-Germany). 
These filters have been available for use by all LBTO partners since semester 2015A.  
However, they must be requested in advance to allow for time to swap with filters normally 
kept in the LBC Red filter wheels.

\begin{table}[ht]
\caption{Overview of Available LBC Filters} 
\label{tab:LBCfilters}
\begin{center}       
\begin{tabular}{|l|l|l|l|l|l|} 
\hline
\rule[-1ex]{0pt}{3.5ex}  {\bf LBC Blue}& {\bf 50$\%$ Cut-On}& {\bf 50$\%$ Cut-Off}& {\bf LBC Red}& {\bf 50$\%$ Cut-On}& {\bf 50$\%$ Cut-Off}\\
\rule[-1ex]{0pt}{3.5ex}                & ($\mu$m)&            ($\mu$m)&                          & ($\mu$m)             &($\mu$m)\\
\hline
\rule[-1ex]{0pt}{3.5ex}   {\it U}-Spectroscopic$^{1}$& 0.332& 0.381&  {\it V}-Bessel&  0.493& 0.577\\
\hline
\rule[-1ex]{0pt}{3.5ex}   {\it U}-Bessel& 0.333& 0.382&  {\it r}-sloan&     0.553&     0.686\\
\hline
\rule[-1ex]{0pt}{3.5ex}   {\it B}-Bessel& 0.375& 0.469&  {\it R}-Bessel&    0.572&     0.690\\
\hline
\rule[-1ex]{0pt}{3.5ex}   {\it g}-sloan&  0.397& 0.550&  {\it i}-sloan&     0.697&     0.836\\
\hline
\rule[-1ex]{0pt}{3.5ex}   {\it V}-Bessel& 0.488& 0.610&  {\it I}-Bessel&    0.713&     0.881\\
\hline
\rule[-1ex]{0pt}{3.5ex}   {\it r}-sloan&  0.552& 0.686&  {\it z}-sloan&     0.830&     ...\\
\hline
\rule[-1ex]{0pt}{3.5ex}   ...&  ...& ...&                {\it Y}&           0.952&     1.110\\
\hline
\rule[-1ex]{0pt}{3.5ex}   ...&  ...& ...&                TiO 784$^{2}$&     0.769&     0.788\\
\hline
\rule[-1ex]{0pt}{3.5ex}   ...&  ...& ...&                CN 817$^{2}$&      0.802&     0.821\\
\hline
\rule[-1ex]{0pt}{3.5ex}   ...&  ...& ...&                F972N20$^{2}$&     0.952&     0.974\\
\hline
\end{tabular}
\end{center}
{\footnotesize $^{1}$ = Broad width (top-hat) filter response designed
to mimic the spectroscopic coverage in this wavelength range; $^{2}$ = Medium width filters}\\
\end{table}

\indent Steps continue toward the improvement of collimation with the LBCs.  The goal
is to expand and improve the range of conditions under which collimation can be achieved. 
Currently, collimation is achieved through a custom IDL program called {\tt DOFPIA} which 
measures and analyzes the pupil (highly de-focused) images of stars, see 
Hill et al. (2008)\cite{2008SPIE.7012E..1MH} for more details.  Using a geometrical method 
described by Wilson (1999) \cite{1999rto..book.....W} aberration coefficients are derived 
by measuring the internal 
and external borders of pupils, as well as in some cases, their illumination profiles.
Empirically determined scaling relations based on these are then used to apply the Zernike
corrections (Z4, Z5, Z6, Z6, Z8, Z11, and Z22) needed to remove the aberrations.  
A small region of Chip 2 is read-out in order to speed up the
process.  The process is repeated until the corrections converge.\\
\indent In September 2015, during fall startup, the position of the region used for 
{\tt DOFPIA} was changed to below the rotator center on Chip 2 for both LBCs
in an effort to improve collimation.  The rotator center for LBC Blue is (in 
detector coordinates on Chip 2) {\tt [1035,2924]} and for LBC Red {\tt [1078,2913]}.
The new section is at {\it Y} = 1201:2608 (previous focus used {\it Y} = 3201:4608)
and the entire width of Chip 2.  {\tt DOFPIA} needs to be run ever 30 minutes or less to 
maintain effective collimation.  In addition, active 
collimation continues during science exposures when both technical chips takes exposures 
every 8-32 seconds (guiding and pupil exposures are set to the same exposure time, which
is dependent on the brightness of the guidestar).  Corrections from the collimation 
technical chip (Tek Chip 2) are applied in between science exposures.  
However, {\tt DOFPIA} does have certain 
limitations that can affect achieving optimal collimation.  This often occurs at the 
start of an observing night, where temperature differentials between the primary mirrors 
and ambient air make collimation time-consuming or difficult to achieve.  The limitations, 
in part, come from issues in how {\tt DOFPIA} fits the extra-focal pupils, as well as the 
signal-to-noise (S/N) of the data.  A new system, Wavefront Reconstruction Software (WRS)
is being developed with INAF.  WRS 
takes into account S/N considerations, applies a new wavefront reconstruction algorithm, 
and maintains a more detailed log of the corrections applied, 
(Stangalini et al. 2014\cite{TechnicalReport...INAF...2014}).  WRS testing indicates
the biggest gains can be made when there is significant coma at the start of the night
(Z7 and/or Z8).  WRS tests have been carried it during 2015 and 2016, but unfortunately
have been hampered by poor weather during engineering nights.  In some cases, WRS tests
have been performed in parallel with science observations, mostly with LBC Red (DX) while 
MODS-1 or LUCI-1 is being used (SX).  It is expected that a final analysis and a beta
version will be made available for more widespread testing in the near future.\\
\indent Improvements to the LBC control hardware and software continue. Software upgrades
are regularly rolled out to correct and improve guiding, instrument rotation, non-sidereal
guiding, and error handling.  As noted in Summers et al. (2104)\cite{2014SPIE.9152E..2ES}, 
upgrades will be made to the LBC control systems since their installation at LBT.  These 
include replacing the four CCD controllers, one each for LBC Blue and Red science CCDs and 
one each for Blue and Red guiding (technical controllers).  Each controller is handled by 
a separate PC running Microsoft Windows Server 2003.  An ethernet CCD controller upgrade 
will replace the need for four physical Windows PCs. Currently a single LINUX machine 
known as the Central Management Unit (CMU) is used to run the LBC software and interface
with the CCD controllers.  A future upgrade will eliminate the need for the Windows 
software and port everything to a LINUX based software system.  A BeagleBones
board ({\url http://beagleboard.org/}) will be used to run LINUX on a daughter card 
attached to the CCD Controller cards.
Step one (completed) is to move image analysis from the Windows PCs to the CMU.  
A Prototype CCD controller with a BeagleBones board is currently
being tested by INAF, as is ongoing porting of software from Windows to LINUX.

\subsection{Multi-Object Double Spectrograph (MODS)}
\label{subsec:MODS}
The Multi-Object Double Spectrographs (MODS) are a pair of identical optical imagers and
spectrographs designed to use longslit and user-designed multi-object slit masks.  Each
MODS is attached to the straight through {\it f/15} Gregorian focus on the respective primary
mirror (MODS-1 on SX, MODS-2 on DX).  MODS were designed and built by The Ohio
State University as part of its contribution to the first generation of LBT instruments. 
Specific details can be found in 
Pogge et al. (2006)\cite{2006SPIE.6269E..0IP},(2010)\cite{2010SPIE.7735E..0AP} and 
first light results are presented in Pogge et al. (2012)\cite{2012SPIE.8446E..0GP}.  The 
instrument description and capabilities described applies to both MODS-1 and MODS-2. 
MODS-1 was installed and aligned in 2009 and became available for partner science in 
semester 2011B. MODS-2 was installed in semester 2014A and commissioned in 2014B-2015.
Both MODS have been used for on-sky science since semester 2015B and have been used
in binocular mode in 2016.\\
\indent MODS is a double spectrograph and imager that employs reflective optics to
achieve high-throughput from near-UV (0.32 $\mu$m) through near-IR (1.1 $\mu$m) 
wavelengths.  Both MODS house separate blue- and red-optimized channels that use
custom-built E2V CCD231-68 back-side illuminated CCDs with 3072 $\times$ 8192 pixels
(15 $\mu$m square).  The blue channel is standard silicon with E2V Astro-Broadband coating
and the red channel is 40 $\mu$m thick deep depletion silicon with extended-red coating
(E2V Astro-ER1).  This provides increased performance long-wards of 0.8 $\mu$m, with 
significantly reduced fringing relative to other optical spectrographs and imagers.
The readout time for the un-binned 8K$\times$3K is $\sim$ 105 seconds.  For more 
information see Atwood et al. (2008)\cite{2008SPIE.7021E..08A}.  \\
\indent The guiding and wavefront-sensing systems are constructed as part of MODS and
located above the instrument focal plane, but within the unit itself.  MODS also houses 
the calibration system internally. It consists of continuum (fixed intensity 
Quartz-Halogen and variable intensity incandescent) used for calibration imaging and 
spectroscopic flats; and emission-line lamps (arc lamps) used for wavelength calibration 
of grating and prism spectroscopy.  The optical
layout of MODS incorporates a dichroic beam splitter below the focal plane that splits
light into separate, but optimized blue and red only channels.  There is a cross-over
at 0.565 $\mu$m that results in a drop in flux in a small region ($\sim$ 0.005 $\mu$m
centered on this wavelength).  For some science cases, users may choose to employ 
blue- or red-only observations. The dichroic is replaced with no optic in the beam for 
blue-only mode and replaced with a flat mirror for red-only mode 
(imaging and spectroscopy).  MODS uses an infrared laser ($\lambda$ = 1.55 $\mu$m)
closed-loop image compensation  system (IMCS) to provide flexure compensation due to 
gravity, mechanical, and temperature effects. The IMCS can null motion to within an
average of $\pm$0.6 pixels for every 15\degrees\  for elevations of 
90\degrees\ -30\degrees. More information about the IMCS can be found in
Marshall et al. (2006)\cite{2006SPIE.6269E..1JM}.\\
\indent MODS has two observing modes:  direct imaging, and spectroscopy using curved
focal plane masks.  These masks include facility longslit and multi-object slit masks
that can be custom designed by users and fabricated at the University Research
Instrumentation Center (URIC) at the University of Arizona
(see Reynolds et al. 2014\cite{2014SPIE.9151E..4BR} for details on fabrication and 
materials used for the masks).
Direct imaging is achieved by replacing the grating with a plane mirror and is used for 
target acquisition for spectroscopy.  The standard acquisition is to read out a 
smaller 1K$\times$1K region of the CCD to reduce overheads during the acquisition 
(readout $\sim$ 40 sec). Direct imaging can also be used for science programs.  MODS 
includes a full complement of sloan filters: {\it u}, and {\it g} for the Blue channel; 
and {\it r}, {\it i}, and {\it z} for the Red channel.  The usable FOV is 
6\arcmin $\times$ 6\arcmin\ but with degraded image quality at radii $>$ 4\arcmin. In the 
case of direct imaging for science, the CCDs are read out in 3K$\times$3K mode (readout 
time is $\sim$ 68 sec).\\
\indent MODS has two spectroscopic modes: a medium resolution diffraction grating
optimized for blue and red spectral regions with {\it R} $\sim$ 2300, and 1850
(using a 0\arcsec.6 wide slit, and scaling inversely with increasing slitwidth), 
respectively; and a double-pass 8\degrees\  glass prism with back reflective coating 
that produces a low-dispersion spectroscopic mode with {\it R} $\sim$ 420-140 in the blue, 
and {\it R} $\sim$ 500-200 in the red.  The grating dispersion
uses the full 8K$\times$3K CCD, while the prism mode uses a 4k$\times$3K readout mode.
Longslit and multi-object slit masks are made available through a mask cassette system
with 24 positions.  Each mask is matched to the shape of the Gregorian focal plane.
The first 12 positions in the cassette contain permanent facility and testing masks.
The facility science masks include:  0\arcsec.3, 
0\arcsec.6, 0\arcsec.8, 1\arcsec.0, and 1\arcsec.2 longslit segmented masks (each
contains five 1\arcmin\ long slits each separated by segmented braces); 
and a 5\arcsec\ wide $\times$ 60\arcsec\ long 
longslit single segment mask used primarily for spectro-photometric calibrations.
In semester 2016A a new facility 2\arcsec.4 $\times$ segmented longslit 
was fabricated and is now available for all LBT partners
upon request.  The remaining 12 mask slots are available for custom designed MOS
masks (discussed in the next section).\\
\indent MODS has excellent sensitivity at both the UV and near-IR extremes, producing 
high S/N spectra as far blue-ward as 0.315 $\mu$m and as far red-ward as 1.05 
$\mu$m.  Table 3 provides an overview of the imaging and spectroscopic modes available 
for MODS-1 and MODS-2.

\begin{table}[ht]
\caption{Overview of MODS-1/-2 Configurations} 
\label{tab:MODSCap}
\begin{center}       
\begin{tabular}{|l|l|l|l|l|l|} 
\hline
\rule[-1ex]{0pt}{3.5ex}  {\bf Mode}&    {\bf Channel}&  {\bf Filters}&      {\bf Resolution}&  {\bf $\lambda$}&  {\bf CCD Size} \\
\rule[-1ex]{0pt}{3.5ex}            &                 &               &      (0\arcsec.6 slit)&  ($\mu$m)&                 \\
\hline
\rule[-1ex]{0pt}{3.5ex}  Imaging&   	Dual&    {\it u},{\it g},{\it r},{\it i},{\it z}&    ...&                        0.33-0.95&          3K$\times$3K\\
\rule[-1ex]{0pt}{3.5ex}         &   	Blue&   {\it u},{\it g}&           					...&                        0.33-0.55&                       \\
\rule[-1ex]{0pt}{3.5ex}         &   	Red &   {\it r},{\it i},{\it z}&   					...&                        0.55-0.95&                       \\
\hline
\rule[-1ex]{0pt}{3.5ex}  Grating&       Dual&   Clear&                                 1850-2300&                        0.31-1.05&         8K$\times$3K\\
\rule[-1ex]{0pt}{3.5ex}  Spectroscopy&  Blue&   Clear&       						   1850 (@0.4$\mu$m)&				0.32-0.60&                     \\
\rule[-1ex]{0pt}{3.5ex}         	&   Red&    GG495&       						   2300 (@0.7$\mu$m)&				0.50-1.05&                     \\
\hline
\rule[-1ex]{0pt}{3.5ex}  Prism&       	Dual&   Clear&                                 500-140&                        	 0.31-1.05&         4K$\times$3K\\
\rule[-1ex]{0pt}{3.5ex}  Spectroscopy&  Blue&   Clear&       						   420-140&                         0.32-0.60&                     \\
\rule[-1ex]{0pt}{3.5ex}              &  Red&    GG495&       						   500-200&                         0.50-1.05&                     \\
\hline
\end{tabular}
\end{center}
\end{table}

Figure 2 shows an example spectrum obtained with MODS-1, using the dual-grating mode
with high sensitivity at both near-UV and near-IR wavelengths.  The target is a 
z$\sim$1 Ultraluminous Infrared Galaxy (ULIRG) that is suspected of being a late-stage 
merger between two gas-rich spiral galaxies 
(Rothberg et al. 2015\cite{2015IAUGA..2257946R}). 
ULIRGs emit 10$^{12}$ {\it L}$_{\odot}$ integrated over 8-100 $\mu$m and contain anywhere 
from 10$^{9}$-10$^{10}$ {\it M}$_{\odot}$ of molecular gas, which provides fuel for 
forming new stars and growing super-massive central black holes (SMBH) that power Active 
Galactic Nuclei (AGN). The most powerful AGN are quasars (QSOs) and reside in massive 
elliptical galaxies (10-100$\times$ more massive than the Milky Way).  In the local 
Universe, ULIRGs are known as the progenitors of QSO host galaxies 
(e.g. Sanders et al. 1988\cite{1988ApJ...325...74S}, 
Rothberg et al. 2013\cite{2013ApJ...767...72R}).

   \begin{figure} [ht]
   \begin{center}
   \begin{tabular}{c} 
   \includegraphics[height=9.5cm]{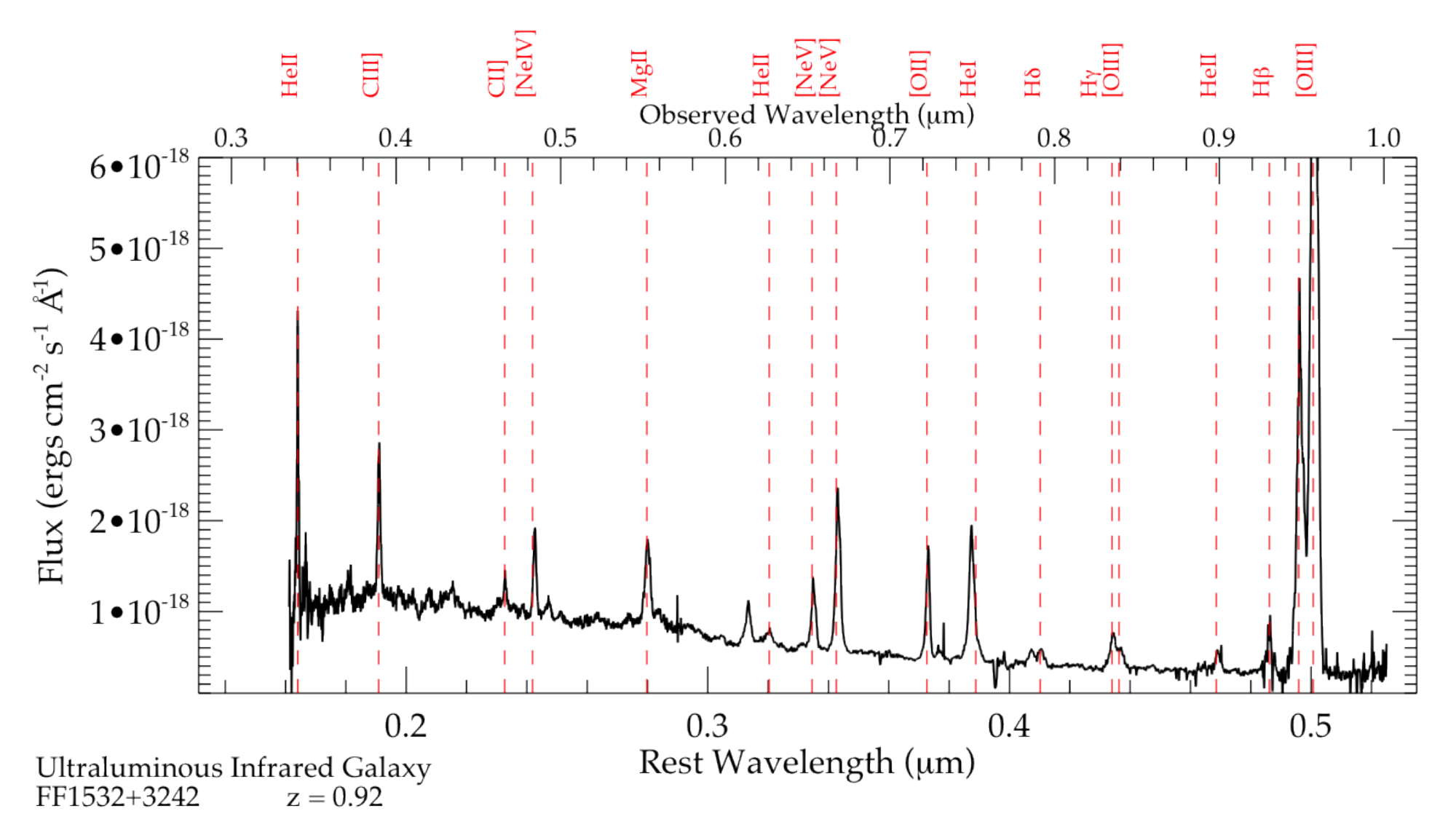}
   \end{tabular}
   \end{center}
   Figure 2:  Shown is a sample spectrum from a project to measure the dynamical, 
   star-formation, and search for AGN in a sample of 
   Ultraluminous Infrared Galaxy mergers at 0.4 $<$ z $<$ 1.0.  The sample selection is
   different than most intermediate redshift ULIRG surveys (PI Rothberg). 
   Rest-frame UV/Optical spectra were obtained from MODS-1/-2 in and used in conjunction 
   with archival imaging from {\it Hubble Space Telescope}. 
   The spectra shown here were obtained with MODS-1, 4800 sec total integration time, 
   0\arcsec.6 slit, and dual grating mode.  Both the observed and rest-frame wavelengths
   are shown to demonstrate the range of MODS and identify the high-excitation emission
   lines.  The z=0.92 ULIRG shows evidence of a SMBH with 
   {\it M}$_{\bullet}$ $\sim$ 10$^{8}$ {\it M}$_{\odot}$, consistent with the SMBH masses
   in nearby Type I AGN, like Mrk 231.
    \end{figure} 

\subsubsection{MODS Multi-Object Spectroscopic (MOS) Masks}
The multi-object spectroscopy (MOS) masks allow PIs to create masks based on the
scientific needs of the targets to be observed.  Masks are designed using a software
program called {\tt MMS} (MODS Mask Simulator).  The software is a modified version of the
LUCI mask software, {\tt LMS} (LUCI Mask Simulator). A user's manual can be found at 
{\url www.astronomy.ohio-state.edu/~martini/mms/}. Both are based upon the 
European Southern Observatory SkyCat tool.  The software allows users to load a {\it fits} 
image file with a valid world coordinate system (WCS) or access archival images from the 
Digital Sky Survey or 2MASS (2 Micron All Sky Survey) and place slits of user-defined 
length and width within the field of view of MODS.   The {\tt MMS} software allows users 
to rotate the image as needed, add multiple slits, and display information
to ensure that slits do not overlap.  Alignment is done using a minimum of three
4\arcsec $\times$ 4\arcsec alignment boxes that are placed over the positions of stars
in the field.  Smaller sized boxes may be used if needed, but they should be larger
than the upper limits of the required seeing constraints (so the star may be fully
measured in the box). The MODS alignment software ({\tt modsAlign}) uses these boxes to 
determine offset and 
rotation to align the mask with the target field.  The newest version of {\tt modsAlign}
used for both MODS auto-detects the alignment boxes using the measured focal-plane
to detector geometries and then prompts users to centroid on the alignment stars that 
should be centered within the box. The more stars used, the more 
precise the alignment.  The software compares the centroid position of the stars and the
positions of the alignment boxes to determine the offsets in translation and rotation.
The {\tt MMS} software also requires users to select a valid guidestar, and provides
an overlay of the Auto-Guiding and Wavefront sensing (AGW) patrol field.  Figure 3 
shows the {\tt MMS} software with an example target and MOS mask being created
({\it left}) and an example of the final mask output as Gerber ({\it gbr}) file.  Masks are
submitted by the partner coordinators to the LBTO Mask Scientist at various mask deadlines
during each semester.  The MOS scientist (currently, B. Rothberg) reviews each mask to 
ensure it meets the criteria of sufficient alignment boxes, no overlapping slits, a 
suitable guidestar, etc.  Approved masks are then sent to URIC for fabrication. For more 
information on fabrication and materials used, 
(Reynolds et al. 2014\cite{2014SPIE.9151E..4BR}). Once 
fabricated they are sent to LBTO and mask IDs are used to catalog the masks and place 
them into inventory for future (and possibly repeated) use.  Mask exchanges are typically 
done just before or during the day of the first partner science block.  Unlike LUCI, the 
mask unit and masks are not cryogenically cooled, allowing each partner science block to 
use all 12 of the available mask slots.  Multiple mask exchanges can be done fairly 
quickly during partner science blocks if more than 12 MOS masks are needed.

   \begin{figure} [ht]
   \begin{center}
   \begin{tabular}{c} 
   \includegraphics[height=5.6cm]{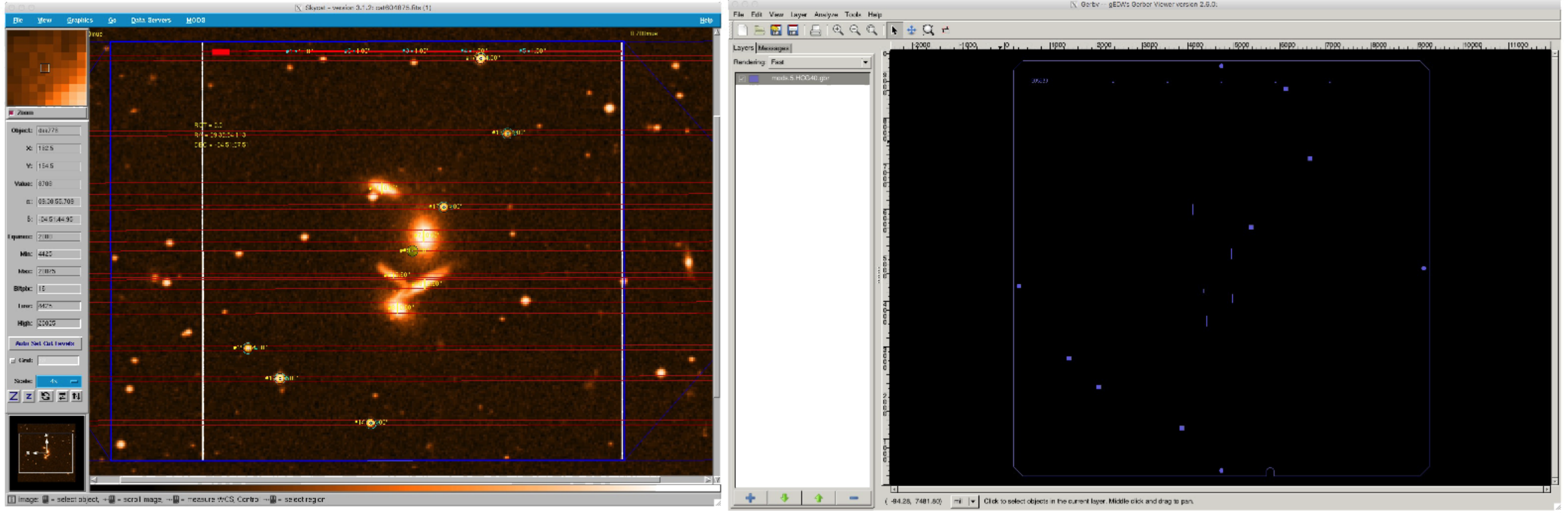}
   \end{tabular}
   \end{center}
   Figure 3:  MODS {\tt MMS} software used to design masks MOS masks ({\it left}).
   The example shown uses Hickson Compact Group 40, a group of five galaxies
   gravitationally bound to each other, several of which are in the early stages
   of interaction.     {\tt MMS} creates a Gerber file ({\it gbr}) which has the 
   information needed to physically manufacture the mask.  Note the square reference 
   boxes used for alignment, and science slits 
   (0\arcsec.6\ width and of 4\arcsec\ - 10\arcsec\ in length).
    \end{figure} 

\subsubsection{Binocular Observing with MODS-1 \& MODS-2}
The use of MODS-1 \& MODS-2 together for scientific observations marks the second of the 
three facility instruments ready for binocular operations on-sky.  The first tests of the
binocular mode of MODS-1 \& MODS-2 took place on UT January 15, 2016.  The instrument PI,
Richard Pogge (Ohio State University) developed an interface that takes a single MODS
script and ``twins'' it so that the preset (or pointing information) instructs the 
binocular mount to move to a designated set of celestial coordinates and configures both 
mirrors to point at the same region of the sky.  The observers simply run a shell script,
for acquisition ({\tt acqBinoMODS}) or for starting science observations 
({\tt execBinoMODS}) which automatically twins the single
input script.  For spectroscopy, after the preset, the observers must then align and place 
the science target in the longslit for each MODS separately.  In the case of imaging,
observers execute a shell script that moves the telescope to the science field and
begins the science integrations on both MODS.\\
\indent Figure 4 shows the first dual grating spectra (1\arcsec.0 slitwidth) obtained 
simultaneously from MODS-1 and MODS-2 of the nearby Seyfert 2 AGN host galaxy NGC 1068. 
The data were processed using a quick-look software developed by The Ohio State University 
(and based upon the modsIDL data reduction package: 
{\url http://www.astronomy.ohio-state.edu/MODS/Software/modsIDL/})
and currently available for visiting astronomers to use to assess the quality of data 
obtained in near real time.  A total of three targets, including an intermediate redshift 
ULIRG, were successfully observed that night using the MODS-Binocular mode.  The 
MODS-Binocular mode has been available on a ``shared-risk'' basis for visiting astronomers 
since May 2016, allowing all LBT partners an opportunity to use both MODS.  
As of semester 2016A, only longslit-longslit or imaging-imaging configurations 
are supported.  MOS masks are not currently supported for MODS-Binocular mode, but should 
be available for use in semester 2016B. This will require two masks to be fabricated from 
a single {\it gbr} file. The next step is to use mixed configurations, such as different 
MOS masks for the same field, or a mixture of spectroscopy and imaging.  Nevertheless, 
MODS-Binocular currently provides PIs a $\sqrt 2$ increase in S/N for the same amount of 
observing time as before.

   \begin{figure} [ht]
   \begin{center}
   \begin{tabular}{c} 
   \includegraphics[height=9cm]{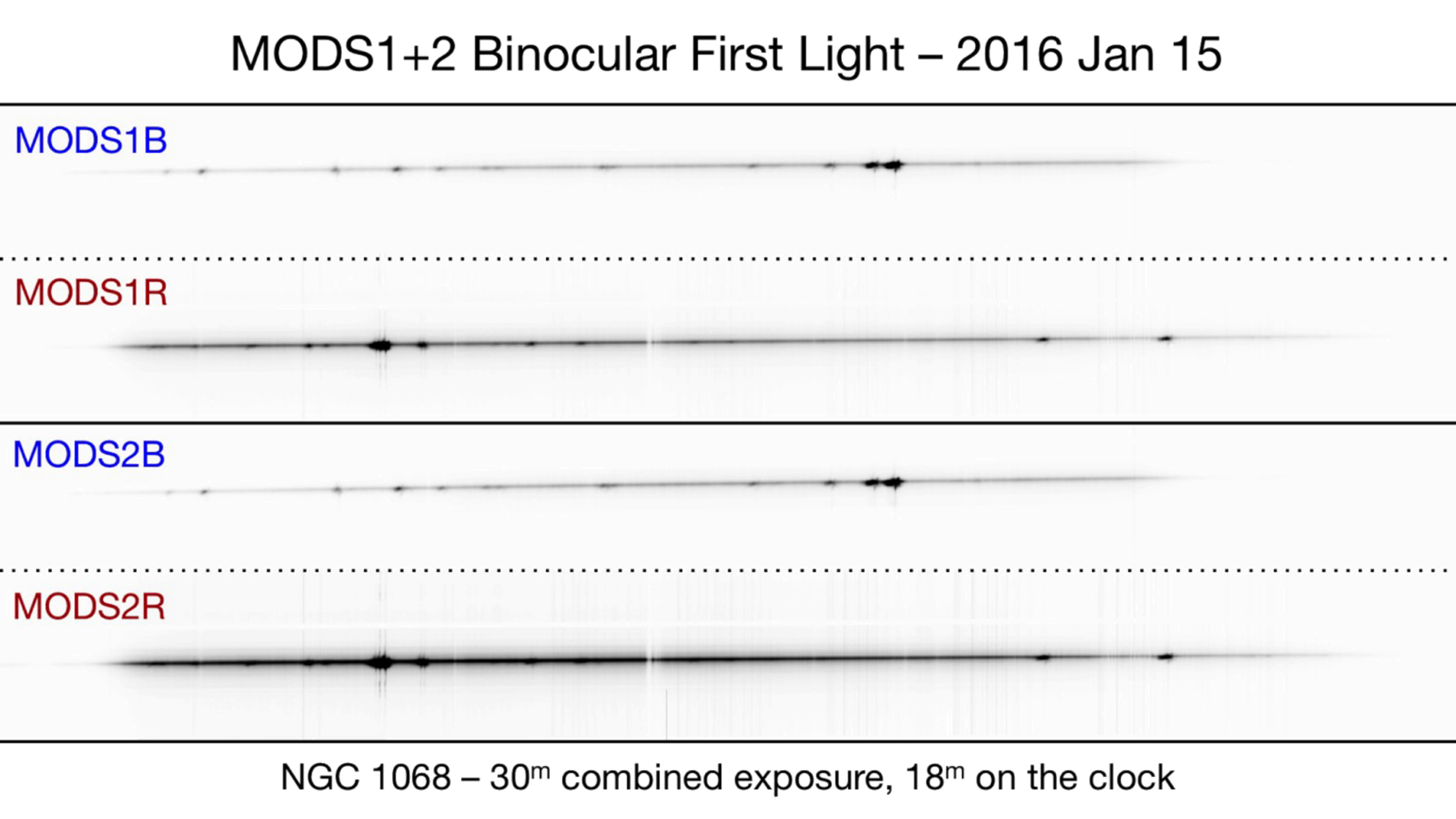}
   \end{tabular}
   \end{center}
   Figure 4:  First MODS-Binocular observations of a single target, the Seyfert 2 AGN
   NGC 1068.  Both MODS were configured in dual grating mode with a 1\arcsec.0 slitwidth.
   The total exposure time was 15 minutes per channel and mirror (300 sec $\times$ 3
   exposures), or 30 minutes total each for blue and red channels.  Thus, including 
   overheads, 30 minutes worth of data were obtained in $\sim$ 18 minutes.  The initial
   binocular acquisition preset was sent by the instrument PI Richard Pogge. The 
   MODS-Binocular observations that night were executed from the remote observing room 
   in Tucson.
    \end{figure} 

\subsection{LBT NIR Spectroscopic Utility with Camera Instruments (LUCI)}
\label{subsec:LUCI}
The two LBT Utility Camera in the Infrared instruments (LUCI, formerly LUCIFER), 
are a pair of cryogenic near-IR (NIR) instruments, capable of imaging and spectroscopy 
(longslit and MOS) each located at a bent Gregorian {\it f/15} focus port of the SX and DX
mirrors.  The discussion of the LUCIs will focus primarily on seeing-limited operations.
Diffraction-limited modes for both LUCIs are still being commissioned.
The LUCIs are rather compact and rely on a series of fold mirrors
to bring the light from the tertiary mirror (M3) into the focal plane.  The LUCIs
are cooled using closed cycle coolers which are monitored to maintain
the correct temperatures needed for optimal operation.  Currently, flexure compensation 
is achieved in a passive mode whereby a lookup-table is used based on the elevation and 
rotation of the instrument and applied before an exposure is taken.  The corrections 
are applied to the last fold mirror in the optical train (FM4), which lies in front of 
the instrument's internal pupil.  An active flexure compensation system is currently in 
development that will apply corrections during a science exposure.\\
\indent The LUCIs are sensitive from 0.95-2.44 $\mu$m and are designed to be used in both 
seeing-limited and diffraction limited (via active optics) modes.  More detailed 
technical information and on-sky performance about LUCI (specifically LUCI-1) can be 
found in Seifert et al. (2003)\cite{2003SPIE.4841..962S},
Ageorges et al. (2012)\cite{2010SPIE.7735E..1LA}, and 
Buschkamp et al. (2012)\cite{2012SPIE.8446E..5LB}. LUCI-1
was installed at LBT in September 2008 and has been in service from December 2009 through 
July 2015 in seeing-limited mode only. It was removed from the telescope during 2015 
summer shutdown to replace the detector with a Hawaii2 RG (H2RG) 2K $\times$ 2K detector 
and install a high resolution camera (N30) which is designed to be used with adaptive 
optics (AO).  These upgrades were designed so LUCI-1 would match the capabilities of 
LUCI-2.  LUCI-2 was made available to the LBT community for on-sky science starting in 
semester 2015B and continuing through semester 2016A in seeing-limited mode only.  
Commissioning of the diffraction-limited modes of LUCI-1 and LUCI-2 are ongoing.  
In addition, both LUCIs are designed to work with the ARGOS, a green laser system 
designed for wide-field ground-layer adaptive optics (GLAO) corrections, 
(e.g. Rabien et al. 2010\cite{2010SPIE.7736E..0ER},
Rabien et al. 2014\cite{2014SPIE.9148E..1BR}, and
Rahmer et al. 2014\cite{2014SPIE.9149E..2AR}).\\
\indent Both LUCI-1 and LUCI-2 are now equipped with the same 2K $\times$ 2K H2RG 
detectors. The detectors are controlled by GEIRS (GEneric InfraRed detector Software)
developed by MPIA.
The LUCIs now have the same set of cameras: an {\it f/1.8} camera with 0\arcsec.25 pixel$^{-1}$ 
(N1.8), an {\it f/3.75} camera with a 0\arcsec.12 pixel$^{-1}$, and an {\it f/30} camera with 
0\arcsec.015 pixel$^{-1}$.  Nominally, the N1.8 camera is primarily used for 
seeing-limited spectroscopy; the N3.75 camera is used for seeing-limited imaging, 
yielding a 4\arcmin $\times$ 4\arcmin\ FOV; and the N30 is used for AO imaging and 
spectroscopy, providing a 30\arcsec $\times$ 30\arcsec\ FOV.  Both LUCIs also 
house the same complement of broad and narrow-band filters.  However, there are 
differences between the available spectroscopic gratings for the two LUCIs.  
Tables 4 \& 5 provide an overview of the capabilities available for both LUCIs in 
seeing-limited mode.  Unlike MODS, where the grating tilt is not changeable by the user, 
the LUCIs offer a wide range of configuration possibilities that can be achieved with 
various tilts (i.e. central wavelengths or $\lambda$$_{\rm c}$), gratings, slits, and 
cameras. Using the N1.8 camera, low resolution grating (G200) permits nearly complete 
coverage of the near-IR window with only two settings.  The high resolution grating 
(G210) with the N1.8 camera allows for nearly full wavelength coverage of each filter 
(i.e. {\it z}, {\it J}, {\it H}, and {\it K}-band).  Users also have the flexibility to
combine cameras, gratings, slits, and $\lambda$$_{\rm c}$ in different ways to achieve
a wide range of scientific goals (i.e. higher spectral resolutions over shorter
wavelength ranges).  

\begin{table}[ht]
\caption{LUCI-1 \& LUCI-2 Filters Available for Science} 
\label{tab:LUCIcap1}
\begin{center}       
\begin{tabular}{|l|l|l|l|l|l|l|} 
\hline
\rule[-1ex]{0pt}{3.5ex}  {\bf Filter}&  $\lambda$$_{C}$&  {\bf FWHM}& &  {\bf Filter}&  $\lambda$$_{\rm c}$&  {\bf FWHM}\\
\rule[-1ex]{0pt}{3.5ex}              &          ($\mu$m)&  ($\mu$m)&  &          &   ($\mu$m)&     ($\mu$m)\\
\hline
\rule[-1ex]{0pt}{3.5ex}        {\it z}&   0.957&            0.195&    &   He I&       1.088&                0.015\\
\hline
\rule[-1ex]{0pt}{3.5ex}        {\it J}&   1.247&            0.305&    & Paschen-$\gamma$& 1.097&            0.010\\
\hline
\rule[-1ex]{0pt}{3.5ex}        {\it H}&   1.653&            0.301&    & OH 1190&       1.194&               0.010\\
\hline
\rule[-1ex]{0pt}{3.5ex}   {\it K}$_{\rm s}$&  2.163&        0.270&    & {\it J} low&   1.199&               0.112\\
\hline
\rule[-1ex]{0pt}{3.5ex}         {\it K}&  2.104&            0.408&    &  Paschen-$\beta$& 1.283&            0.012\\
\hline
\rule[-1ex]{0pt}{3.5ex}    {\it zJ} spec&  1.175&            0.405&    &  {\it J} high&  1.303&              0.108\\
\hline
\rule[-1ex]{0pt}{3.5ex}    {\it HK} spec&  1.950&            0.981&    &  FeII&          1.646&              0.018\\
\hline
\rule[-1ex]{0pt}{3.5ex}        {\it Y}1&  1.007&            0.069&    &  H$_{\rm 2}$&   2.124&              0.023\\
\hline
\rule[-1ex]{0pt}{3.5ex}         OH 1060&  1.065&            0.010&    &  Brackett-$\gamma$& 2.170&          0.024\\
\hline
\rule[-1ex]{0pt}{3.5ex}        {\it Y}2&  1.074&            0.065&    &  ...&             ....&             ...\\
\hline
\end{tabular}
\end{center}
\end{table}

\begin{table}[ht]
\caption{LUCI-1 \& LUCI-2 Orders, Gratings, Valid $\lambda$$_{\rm c}$, and Resolution Values for Seeing-Limited Mode} 
\label{tab:LUCIcap1}
\begin{center}       
\begin{tabular}{|l|l|l|l|l|l|l|} 
\hline
\rule[-1ex]{0pt}{3.5ex}  {\bf Order}&  {\bf G210 HiRes}&   $\Delta$$\lambda$&  Resolution&  {\bf G200 LoRes}& $\Delta$$\lambda$& Resolution\\
\rule[-1ex]{0pt}{3.5ex}             &  Valid $\lambda$$_{\rm c}$ ($\mu$m)&   ($\mu$m)&  (0\arcsec.5\ slit)& Valid $\lambda$$_{\rm c}$ ($\mu$m)&        ($\mu$m)& (0\arcsec.5\ slit)\\
\hline
\rule[-1ex]{0pt}{3.5ex}    1&           ...&              ...&  ...&  1.476-2.535 (L1)&  0.880&  1900 (H), 2600 (K)\\     
\rule[-1ex]{0pt}{3.5ex}     &           ...&              ...&  ...&  1.248-2.491 (L2)&  ''&            ''\\     
\hline
\rule[-1ex]{0pt}{3.5ex}    2&           2.098-2.728 (L1)&  0.328&  5000& 0.897-1.358$^{*}$ (L1)&  0.440&  2100 (z), 2400 (J) \\
\rule[-1ex]{0pt}{3.5ex}     &           2.061-2.702 (L2)&   ''  &    ''& 0.599-1.246$^{*}$ (L2)&  ''&             ''\\
\hline
\rule[-1ex]{0pt}{3.5ex}    3&           1.398-1.819 (L1)&  0.202&  5900&  ...&        ...& ...\\
\rule[-1ex]{0pt}{3.5ex}     &           1.374-1.801 (L2)&    '' &    ''&  ...&        ...& ...\\
\hline
\rule[-1ex]{0pt}{3.5ex}    4&           1.084-1.364 (L1)&  0.150&  5800&  ...&         ...& ...\\
\rule[-1ex]{0pt}{3.5ex}     &           1.072-1.351 (L2)&    '' &   ''&   ...&         ...& ...\\
\hline
\rule[-1ex]{0pt}{3.5ex}    5&           0.839-1.075$^{*}$ (L1)&  0.124&  5400&  ...&        ...& ...\\
\rule[-1ex]{0pt}{3.5ex}     &           0.824-1.066$^{*}$ (L2)&    '' &    ''&   ...&        ...& ... \\
\hline
\end{tabular}
\end{center}
{\footnotesize L1 is LUCI-1, L2 is LUCI-2. * $=$ While $\lambda$$_{\rm c}$ $<$ 0.95 $\mu$m are
valid, the L1 and L2 entrance windows are now coated to cutoff at $\lambda$ $<$ 0.95 $\mu$m.  
Resolution scales down as slitwidth increases.  The {\it zJ} spec and {\it HK} spec 
filters in Table 4 are primarily used with the G200 
grating. The G150 Ks grating is only available on LUCI-1 and the allowable $\lambda$$_{\rm c}$
wavelength range is 1.95-2.4 $\mu$m and $\Delta$$\lambda$ $=$ 0.533 $\mu$m with {\it R} 
$\sim$ 4150 using a 2 pixel slitwidth (0\arcsec.5\ ). The G040 AO grating is only 
available on LUCI-2. More information regarding the G040 AO grating will be determined 
at a later date.  The wavelength coverage in this table assumes the N1.80 camera.  
If using the N3.75 camera 
multiply $\Delta$$\lambda$ by 0.48, and if using the N30 camera multiple by 0.06. 
This table should allow one to determine the wavelength range visible for different 
configurations.}\\
\end{table}

\indent The calibration unit for both LUCIs are external the instrument.  The units are 
mounted above the bent Gregorian ports housing LUCI.  In stowed position, they are flush
with the ports. When required, they are activated by scripts (or manually from the user
interface or WEB/IO interface) and swing out laterally and then downwards so they are 
directly in front of the entrance window.  Three Halogen lamps of varying brightness are 
used for imaging and spectroscopic flats.  In 2016A, a neutral density filter
was added to the LUCI-2 calibration unit to better match the intensities of the LUCI-1 
calibration unit.
Three emission-lamps (arc lamps) are available for wavelength calibration:  Neon, 
Argon, and Xenon.

\subsubsection{LUCI Multi-Object Spectroscopic (MOS) Masks}
\indent LUCI-1 and LUCI-2 each use a cryogenic MOS unit to house both a set of permanent
facility longslit masks and user designed MOS slit masks 
(Hofmann et al. 2004\cite{2004SPIE.5492.1243H} and 
Buschkamp et al. 2010\cite{2010SPIE.7735E..79B}).
The MOS units hold up to 33 masks distributed over two cabinets.
The permanent cabinets, which are inside the LUCIs, houses 10 facility masks, including
longslit masks. These include:  a wide mask with two long slits, one 2\arcsec\ wide (top)
and one 1\arcsec.5\ wide (bottom), both 100\arcsec in length; and longslits of length
3\arcmin.8 and widths of 1\arcsec.0, 0\arcsec.75\ (currently only available in LUCI-2), 
0\arcsec.5, 0\arcsec.25, and 0\arcsec.13 (to be used with AO and the N30 camera).
This main unit houses the focal plane unit (FPU) which places the masks in and out of 
the LUCI focal plane using a robotic grabber arm (see Figure 5).  The grabber slides 
along set of rails to select the requested mask, place it in the FPU, and later place the 
mask back in its designated slot once it is no longer needed (and another mask is 
requested).  When imaging mode is used, an empty mask holder is placed in the FPU to allow 
light to pass unobstructed to the detector. A second exchangeable cassette which contains 
23 masks slots is used to house MOS masks custom designed by science PIs. Secondary 
cabinet exchanges are executed on a monthly basis to accomodate different partner
science programs.  The exchanges include masks from multiple partners, with each
partner assigned a maximum number available slots in the secondary cabinet. Mask exchanges
are performed at cryogenic temperatures and require the use of two 
auxiliary cryostats in order to maintain pressure and temperatures at all times.
An auxiliary cryostat holding a secondary cabinet is loaded with the next set of masks to 
be used for science.  It is evacuated and cooled over 24-48 hours before 
a scheduled exchange. During the exchange, one aux cryostat is attached to LUCI using a 
set of gate valves controlled by software. Rails connect the aux cryostat to LUCI. 
The current installed secondary cabinet is moved along the rails into the cryostat.
That cryostat is removed and a second cryostat is then attached and a secondary cabinet
containing the new masks is placed into LUCI.  The cabinet exchange is all done on the
telescope infrastructure itself.  This requires the cryostats to be lifted up through 
large doors in the high bay up and over the telescope and then gently placed on
a platform on the telescope (located between the SX and DX mirrors where the bent 
Gregorian foci are located)

   \begin{figure} [ht]
   \begin{center}
   \begin{tabular}{c} 
   \includegraphics[height=9cm]{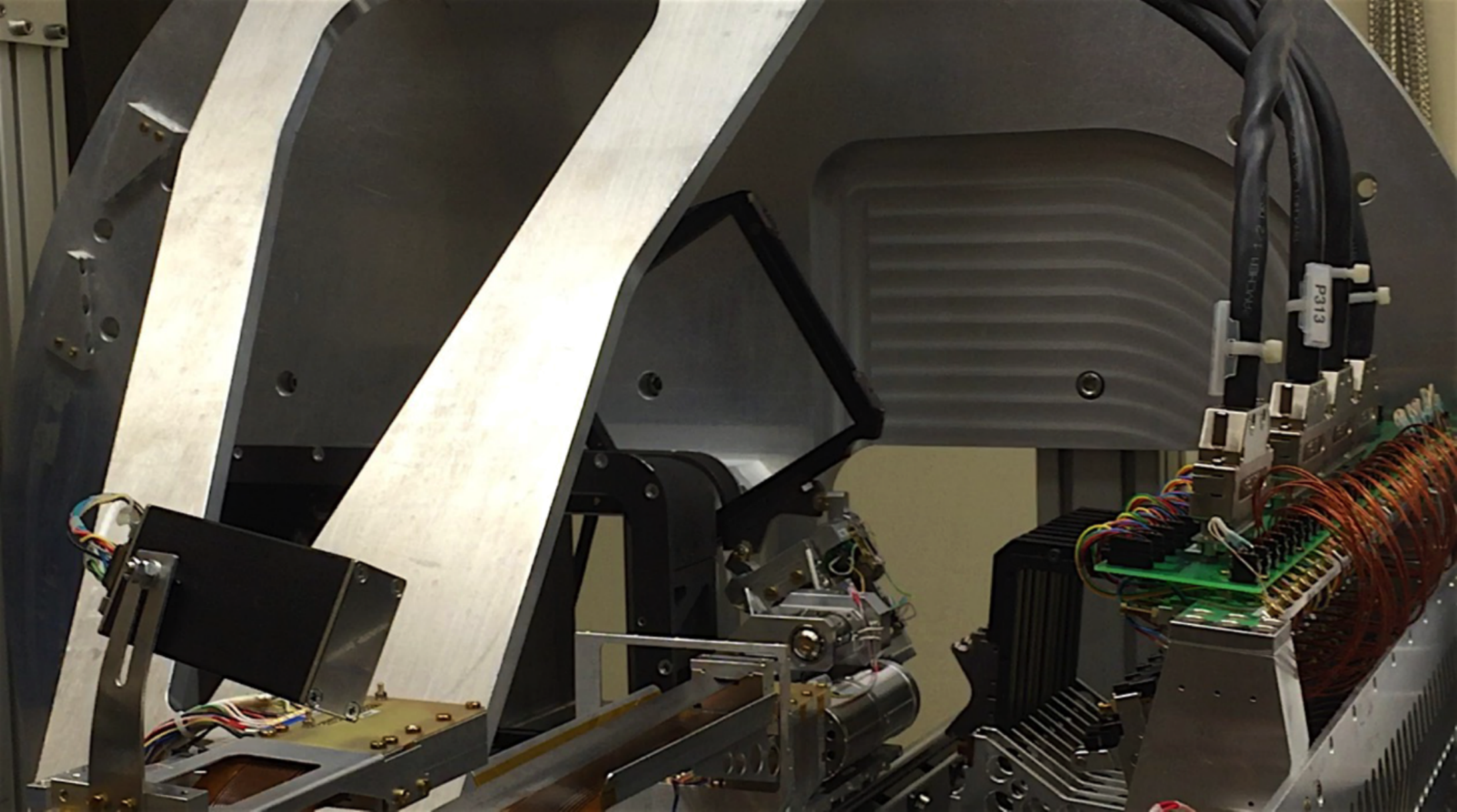}
   \end{tabular}
   \end{center}
   Figure 5:  The LUCI-1 MOS unit outside of its housing and only with the permanent
   cabinet.  Shown is the grabber arm placing a mask holder (which would normally contain 
   a MOS or longslit mask) into the FPU. The rail system can be seen at the bottom of the
   image along with the slots holding other masks in place (right side of photo). Photo
   courtesy of B. Rothberg.
    \end{figure} 

\indent The {\tt LMS} software is used to create MOS mask designs 
({\url http://abell.as.arizona.edu/\~{}lbtsci/Instruments/LUCIFER/})
It is the forerunner of {\tt MMS}.  The interface and concept is 
nearly the same as {\tt MMS} (see Figure 3).  The main differences are the the different 
guidestar patrol fields (generally above the science FOV for LUCI and generally below the 
science FOV for MODS), and that LUCI uses reference stars to calculate the rotation and 
shift needed to align the mask correctly.  A previous version of the LUCI control software 
only required alignment stars to be defined by {\tt LMS} without the needs for alignment 
boxes to be cut in the mask around each star.  Currently, the new version of the LUCI 
software requires both designated reference stars and alignment boxes cut around them, but
future updates may return to a system that does not require physical alignment boxes
cut into the mask. MOS mask designs are submitted at the same deadline as MODS MOS mask 
designs.  As with MODS, the MOS scientist reviews each mask to ensure it meets the 
criteria of sufficient alignment boxes, no overlapping slits, a suitable guidestar, etc.
Approved masks are then sent to URIC for fabrication. For more information on fabrication
and materials used, see Reynolds et al. 2014\cite{2014SPIE.9151E..4BR}. Once fabricated 
they are sent to LBTO and mask IDs are used to catalog the masks and place them into 
inventory for future (and possibly repeated) use. 

\subsubsection{LUCI Software Upgrade}
As noted above, a series of hardware upgrades were made to LUCI-1 during summer shutdown
of 2015. In addition to hardware upgrades, a new LUCI User interface has been developed
by MPIA.  The software completely replaces the previous version used to run LUCI-1.
The new LUCI User interface communicates directly with both LUCI-1 and LUCI-2, and in
principle, allows for binocular control of both instruments.  The interface has three
major components (see Figures 6 and 7):  1)  The Main Observer Graphical User Interface (GUI) 
which displays information such as target coordinates, name, offsets, instrument 
configuration, and integration times in the central ``queue'' panel. Users load the 
script and click ''GO'' which sends the preset information to the telescope control 
software (TCS) and configures the instrument.  Users may Pause, Reset, Abort, and Skip 
(or skip to) steps; 2) the Real-Time Display (RTD), which is based on the ALADIN software 
({\url http://aladin.u-strasbg.fr/}) and is used for longslit and MOS mask acquisition; 
and 3) the Readout and Instrument Control panels, which gives users the ability to 
manually change parameters such as filters, mask, integration times, cameras, and the 
calibration unit.  The previous LUCI-1 software parsed scripts using ASCII plain text.  
The new LUCI control software requires observing scripts in XML format.  Currently,
a scripting webpage, SCRIPTOR ({\url http://scriptor.tucson.lbto.org/}), 
was created to allow users to generate XML scripts by selecting the instrument, 
its configuration, integration times, guidestars, etc.

   \begin{figure} [ht]
   \begin{center}
   \begin{tabular}{c} 
   \includegraphics[height=12cm]{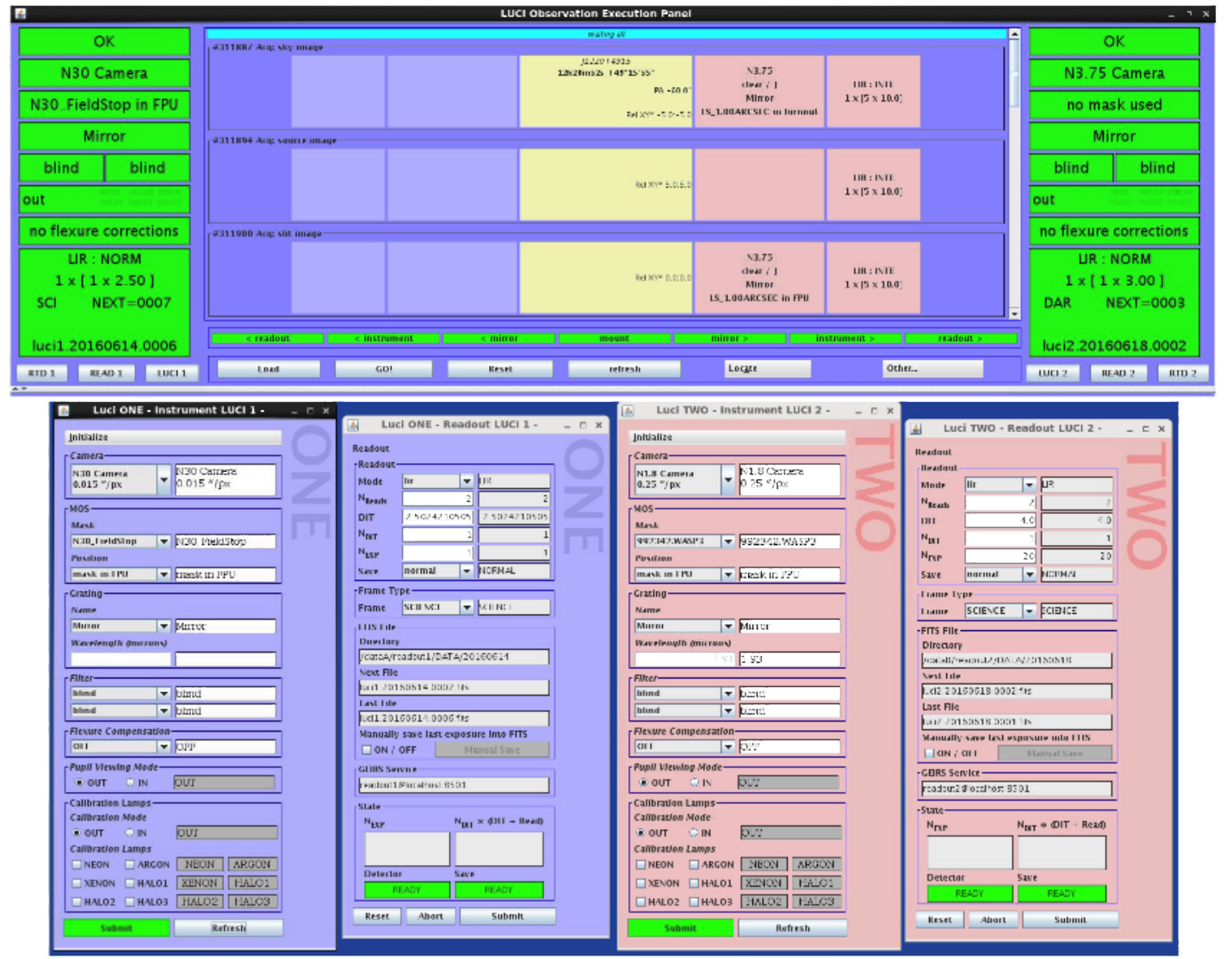}
   \end{tabular}
   \end{center}
   Figure 6:  The new LUCI software interface.  ({\it Top}) The main Observer GUI
   where scripts are loaded and information about the status of the observations
   are presented.  ({\it Bottom}) Readout and Instrument panels for both LUCIs
   which give the users more manual control over the instrument.
    \end{figure} 

   \begin{figure} [ht]
   \begin{center}
   \begin{tabular}{c} 
   \includegraphics[height=10cm]{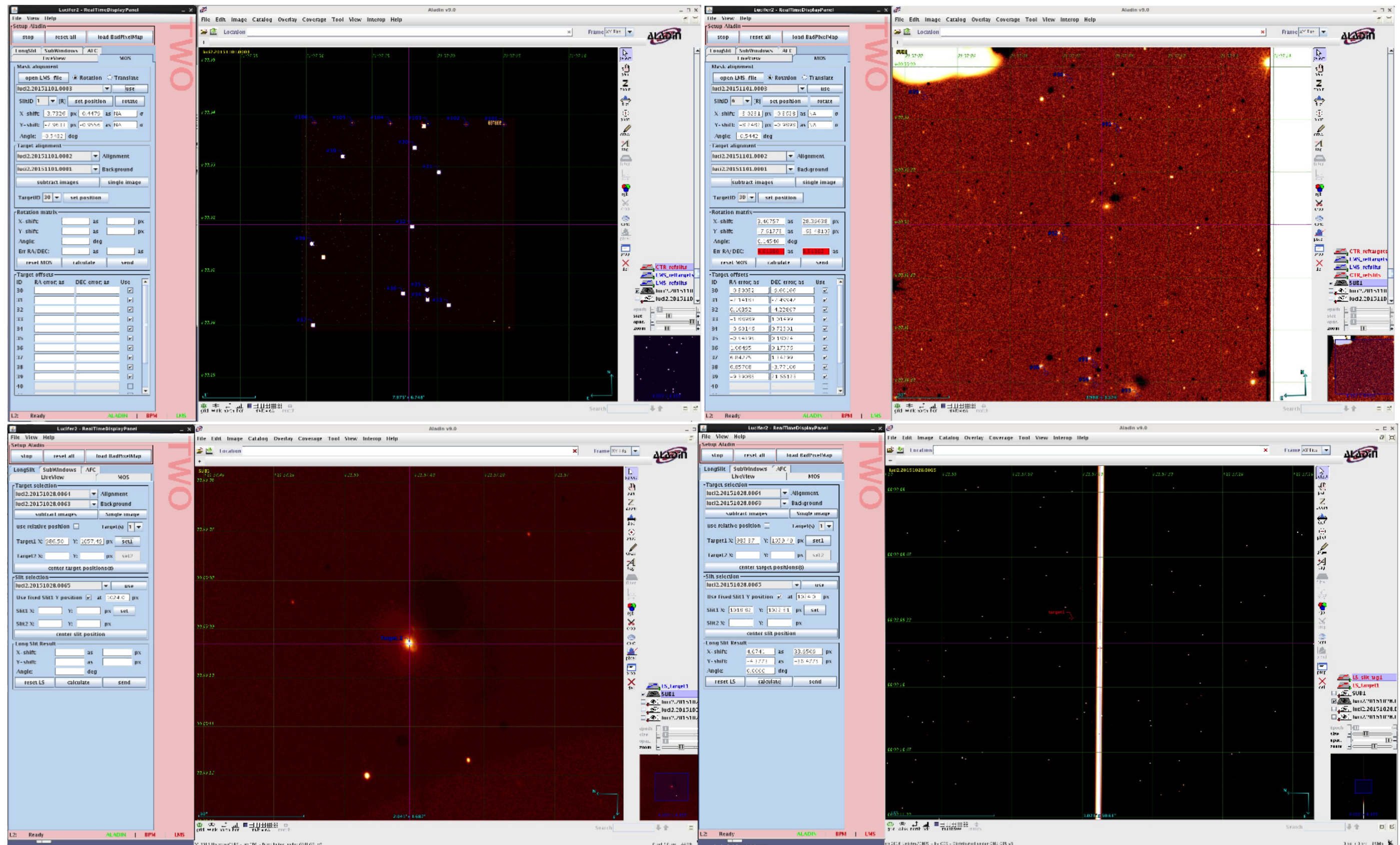}
   \end{tabular}
   \end{center}
   Figure 7:  The Real Time Display showing example of MOS mask acquisition ({\it Top})
   and longslit acquisition ({\it Bottom}).
    \end{figure} 

\section{Mixed-Mode Use}
\label{subsec:mixed}
The goal of LBT is to use the telescope in binocular mode all of the time.  LBTI and LBCs
have been observing in binocular mode for some time.  In the last few months, MODS have 
been successfully tested and used on-sky in binocular mode.  While the 
facility instruments have been designed to work in pairs in binocular mode, the telescope
can also be figured to use instruments in a ``mixed mode.'' These modes would see 
configurations such as MODS-1/LBC-R, MODS-2/LBC-B, LUCI-1/LBC-R, LUCI-2/LBC-B, 
or MODS/LUCI.  Mixed-Mode use is desirable as it opens up a much wider wavelength
ranges for scientific study (i.e. simultaneous UV and near-IR observations).  
As noted in Hill et al. (2014)\cite{2014SPIE.9145E..02H}, the two sides are not required
to have precisely the same target or position angle for binocular mode to work.
The telescope mount points near the the mid-point between the two sides and the telescope
software ``knows'' to avoid presets or small offsets that would violate the co-pointing 
limit (the maximum travel distance between the two mirrors).  Currently, the co-pointing
limit is set by software to be 40\arcsec\ (radius) apart in any direction. In effect, 
once the telescope mount has slewed to a set of coordinates, the two mirrors can 
effectively be treated as independent telescopes.  Each side can dither as required by 
the science, so long as the two sides together don't violate the co-pointing limit.\\
\indent However, a current limitation of using Mixed-Mode is the ability to pass 
a binocular preset from two different instruments to the TCS.  Since 2014, several
combinations of Mixed-Mode have been used.  Currently, different combinations require 
somewhat different setups and have different limitations.  The first attempts in 2013
used a LUCI-1/LBC-R and MODS-1/LBC-R.  In the case of the former, the telescope is
configured in binocular mode, and the TCS waits to receive a preset sent from
each instrument before moving to the field. Once there, LUCI-1 acquisitions are done 
normally (now using the RTD interface) and the script can dither as needed
by near-IR observations, while LBC-R can either stare or dither.  
In the case of MODS-1/LBC-R, the telescope is set up in a hybrid configuration called 
``pseudo-monocular.''  MODS-1 ``drives'' the mount, i.e. the preset is sent only by MODS-1
while LBC-R is ``along for the ride.'' On the MODS side, the acquisition is sent
normally using the {\tt acqMODS} Perl script.  Once the preset is successful, and the 
telescope is at the target field, the LBC-R is collimated using {\tt DOFPIA} and then
a modified LBC script is executed that includes a value of -90$^{\circ}$ in the 
Declination coordinate, which is interpreted by the TCS as a flag to ignore the preset.
Alignment and acquisition proceed as normal with {\tt modsAlign} and the observations
are started using the Perl script {\tt execMODS}.
Both sides can guide independently.  Dithering can be done using the primary mirrors.
Thus, both the dominant (in this case MODS) and the passive side (in this case LBC-R)
can dither independently via the mirror.   In normal LBC binocular mode,
any dithering is done by the mount (which is faster), and not by the two mirrors.
However, it was found that in pseudo-monocular mode when LBC (either Red or Blue)  
dithers using the mirror (which is slower than the mount), the start of the exposure 
does not wait for the slower move and collimation update from the primary mirror. 
Work to correct this is underway.
Figure 8 plots the limits of the M1 (primary) and M2 (secondary) on SX (configured
with MODS-1) with M1 on DX configured with LBC Red.  The plot shows the co-pointing
limits and how they are affected by the motions available to M1 and M2 together.

   \begin{figure} [ht]
   \begin{center}
   \begin{tabular}{c} 
   \includegraphics[height=11cm]{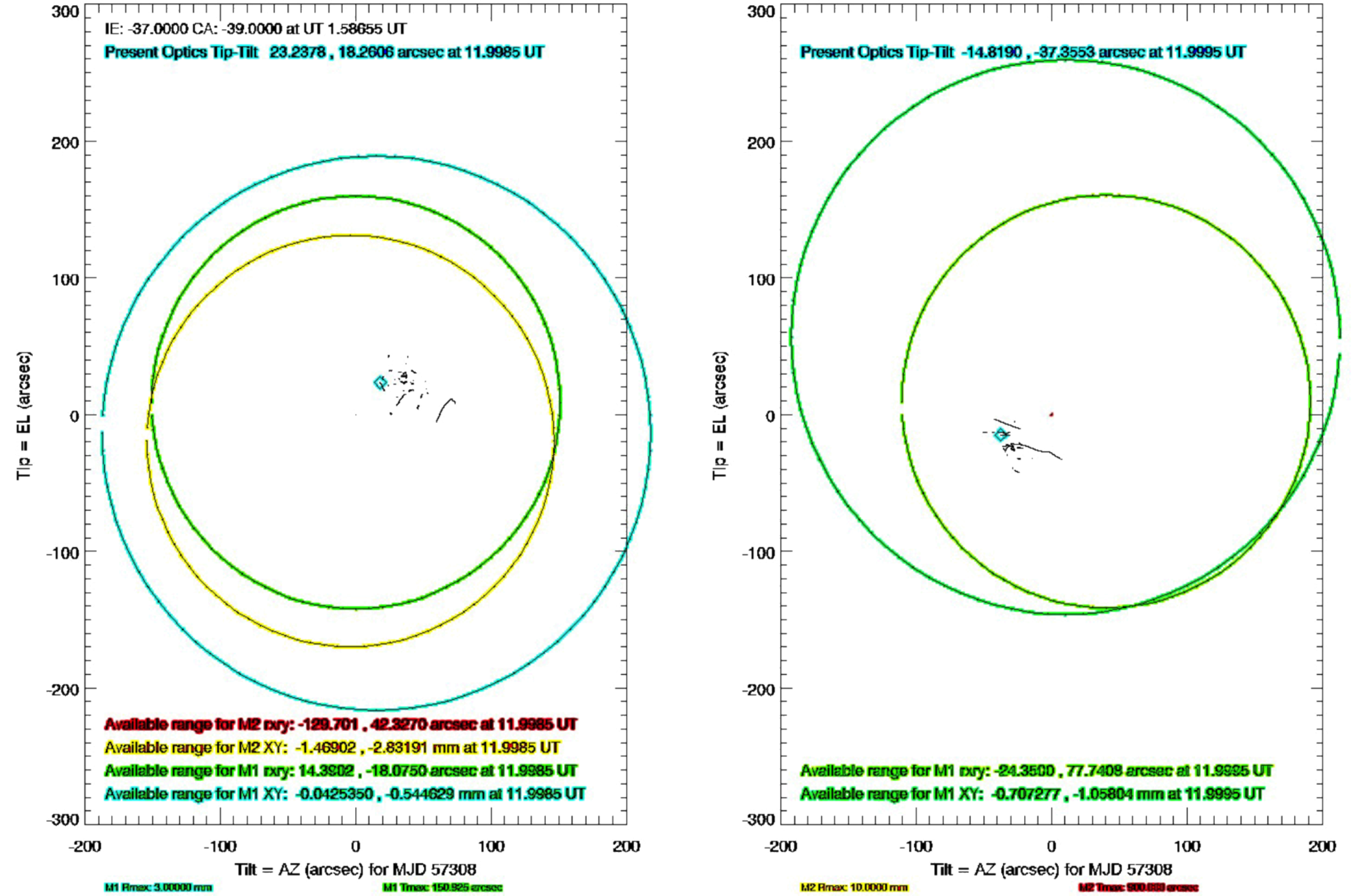}
   \end{tabular}
   \end{center}
   Figure 8:  ({\it Left}) The co-pointing limits plot for MODS-1 (SX) and ({\it Right})
   LBC Red (DX) with the telescope configured in pseudo-monocular mode averaged over one 
   hour.  The circles indicate the various available ranges of re-pointing available to 
   each side of the telescope.  Since LBC Red is at prime focus only M1 is shown on the 
   {\it Right}.  The small blue diamond represents the current pointing.  The XY range 
   of M2 and the tip-tilt range of M1 provide the tightest constraints.  The constraints
   shown here must be considered by PIs in designing their science programs when using 
   Mixed-Mode or Binocular Mode with LUCI or MODS (LBCs dither by moving the mount only, 
   while in Mixed-Mode LBC dithers are done by M1).  
    \end{figure}

\indent With LUCI-1 off the telescope, and with MODS-1 unavailable in November 2015,
Mixed-Mode has been successfully tested with MODS-2/LBC-B and LUCI-2/LBC-B.
As of this writing, LBTO has not tested a Mixed-Mode LUCI/MODS combination.
Based on current limitations, it is theoretically possible this mode should work
in ``pseudo-monocular'' mode.  Since LUCI observations (imaging and spectroscopy)
require dithering, the most effective combination is LUCI as the dominant instrument
and MODS in passive mode.  This mode is scheduled to be tested in the near future.
 
\section{Non-Sidereal Guiding}
Non-sidereal guiding for all three facility instruments is accomplished using the NSIGUI 
(Non-Sidereal Instrument Graphical User Interface).  Figure 9 shows the NSIGUI panels
and all three tabs (left to right).  The simplest method is to either load a properly
formatted ephemeris, or to search for the non-sidereal target and retrieve the ephemeris
using the JPL HORIZONS database 
(web interface: {\url http://ssd.jpl.nasa.gov/horizons.cgi}).  In cases where an ephemeris
does not exist, the middle tab can be used to enter the information on UT time, 
celestial coordinates, and rates manually.  The user then selects the ``SET'' button
in the IIF Non-Sidereal Override Control which sets an override flag in the TCS.
Any preset that is sent from LBC, LUCI, or MODS is then overriden or ``hijacked''
by the non-sidereal coordinates from the NSIGUI.  Instrument configurations are not 
affected by the GUI.  Guiding and tracking then proceeds at a non-sidereal rate.  
The third tab allows users to update the rates as might be needed during observations.
Non-sidereal observations using the LBCs have been done on a regular basis for several 
semesters.  Non-sidereal guiding with LUCI and MODS have been tested at rates up to 
$\sim$ 100\arcsec\ hour$^{-1}$, thus making it possible to obtain spectra of
non-sidereal targets.

   \begin{figure} [ht]
   \begin{center}
   \begin{tabular}{c} 
   \includegraphics[height=11.5cm]{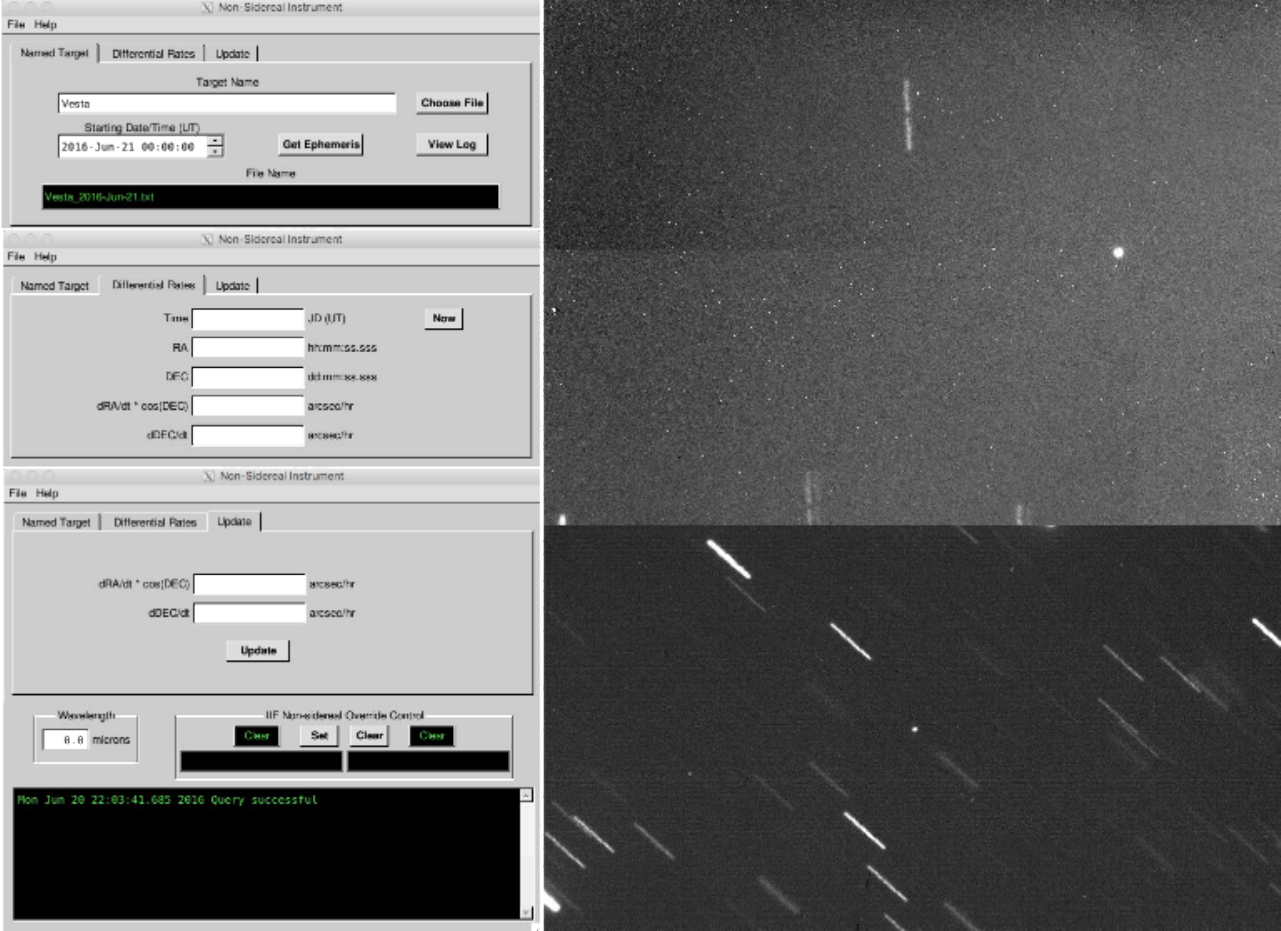}
   \end{tabular}
   \end{center}
   Figure 9:  ({\it Left}) The three tabs for the NSIGUI  (Non-Sidereal Instrument
   Graphical User Interface).  This GUI is used to ``hijack'' preset coordinates from
   the facility instruments and replace them with the coordinates determined from an 
   ephemeris or input along with rates in Tab 2.  ({\it Top Right}) Example of 
   non-sidereal guiding with LUCI-1 of the asteroid Janesick (106\arcsec\ hour$^{-1}$,
   300 sec total time on target - 20 $\times$ 15 sec exposures at {\it z}-band)
   obtained by O. Kuhn.  ({\it Bottom Right}) LBC Red image of the near-Earth asteroid
   2012 ER3 (100\arcsec\ hour$^{-1}$, 100 seconds at Sloan {\it r}) 
   from Hill et al. (2012)\cite{2012SPIE.8444E..1AH}.
    \end{figure}

\section{Summary}
\label{sec:concluson}
Although it was in 2014 that all of the facility instruments arrived and were installed 
on the telescope, it has been 2015 and 2016A that have been truly exciting.  For
the first time, all facility instruments have been used on-sky to obtain scientific
data.  Two-thirds of the facility instruments are available and have been used
in full binocular mode, and it is expected that full binocular observations will occur
for {\it all} facility instruments in 2016B.  Full binocular observations will be
a paradigm shift for the nightly operations of LBTO.  The parameter space of what LBT
can do will significantly expand with full binocular mode, especially in the case 
of Mixed-Mode observations.  Planning tools and scripting tools to address this 
are already in development and testing, led by the development of queue observing at LBT
(see Edwards et al. in the companion proceedings 9910,  Observatory Operations:
Strategies, Processes, and Systems VI).  Full binocular mode will also require a shift
in how users propose and design observations to take full advantage of the capabilities 
of LBT.  Information regarding the design and use of instrumentation can be found 
on the Science Operations webpages: {\url http://scienceops.lbto.org/sciops\_cookbook/}
and {\url http://abell.as.arizona.edu/\~{}lbtsci/scihome.html} as well as the main LBT
webpage: {\url www.lbto.org} which contains the latest updates and information on
all that is happening at the observatory.

\acknowledgments 
 B.Rothberg would like to acknowledge a NASA Keck PI Data Award (Contract $\#$1462408), 
 administered by JPL and the NASA Exoplanet Science Institute, for support of the work on 
 intermediate redshift ULIRGs.



\begin{thebibliography}{45}

\bibitem{2004SPIE.5489..603H}
{Hill}, J.~M. and {Salinari}, P., ``{The Large Binocular Telescope project},''
  in [{\em Ground-based Telescopes}{\nolinebreak\hspace{0.1em}]},  {Oschmann},
  Jr., J.~M., ed., {\em Proc. SPIE} {\bf 5489},  603--614 (Oct. 2004).

\bibitem{2006SPIE.6274E..23A}
{Ashby}, D.~S., {McKenna}, D., {Brynnel}, J.~G., {Sargent}, T., {Cox}, D.,
  {Little}, J., {Powell}, K., and {Holmberg}, G., ``{The Large Binocular
  Telescope mount control system architecture},'' in [{\em Society of
  Photo-Optical Instrumentation Engineers (SPIE) Conference
  Series}{\nolinebreak\hspace{0.1em}]},  {\em Proc. SPIE} {\bf 6274},  627423
  (June 2006).

\bibitem{2006SPIE.6267E..0YH}
{Hill}, J.~M., {Green}, R.~F., and {Slagle}, J.~H., ``{The Large Binocular
  Telescope},'' in [{\em Society of Photo-Optical Instrumentation Engineers
  (SPIE) Conference Series}{\nolinebreak\hspace{0.1em}]},  {\em Proc. SPIE}
  {\bf 6267},  62670Y (June 2006).

\bibitem{2010SPIE.7733E..0CH}
{Hill}, J.~M., {Green}, R.~F., {Ashby}, D.~S., {Brynnel}, J.~G., {Cushing},
  N.~J., {Little}, J., {Slagle}, J.~H., and {Wagner}, R.~M., ``{The Large
  Binocular Telescope},'' in [{\em Ground-based and Airborne Telescopes
  III}{\nolinebreak\hspace{0.1em}]},  {\em Proc. SPIE} {\bf 7733},  77330C
  (July 2010).

\bibitem{2014SPIE.9147E..05W}
{Wagner}, R.~M., {Edwards}, M.~L., {Kuhn}, O., {Thompson}, D., and {Veillet},
  C., ``{An overview and the current status of instrumentation at the Large
  Binocular Telescope Observatory},'' in [{\em Ground-based and Airborne
  Instrumentation for Astronomy V}{\nolinebreak\hspace{0.1em}]},  {\em Proc.
  SPIE} {\bf 9147},  914705 (July 2014).

\bibitem{2008SPIE.7014E..0NS}
{Strassmeier}, K.~G., {Woche}, M., {Ilyin}, I., {Popow}, E., {Bauer}, S.-M.,
  {Dionies}, F., {Fechner}, T., {Weber}, M., {Hofmann}, A., {Storm}, J.,
  {Materne}, R., {Bittner}, W., {Bartus}, J., {Granzer}, T., {Denker}, C.,
  {Carroll}, T., {Kopf}, M., {DiVarano}, I., {Beckert}, E., and {Lesser}, M.,
  ``{PEPSI: the Potsdam Echelle Polarimetric and Spectroscopic Instrument for
  the LBT},'' in [{\em Ground-based and Airborne Instrumentation for Astronomy
  II}{\nolinebreak\hspace{0.1em}]},  {\em Proc. SPIE} {\bf 7014},  70140N (July
  2008).

\bibitem{2008SPIE.7013E..28H}
{Hinz}, P.~M., {Bippert-Plymate}, T., {Breuninger}, A., {Connors}, T., {Duffy},
  B., {Esposito}, S., {Hoffmann}, W., {Kim}, J., {Kraus}, J., {McMahon}, T.,
  {Montoya}, M., {Nash}, R., {Durney}, O., {Solheid}, E., {Tozzi}, A., and
  {Vaitheeswaran}, V., ``{Status of the LBT interferometer},'' in [{\em Optical
  and Infrared Interferometry}{\nolinebreak\hspace{0.1em}]},  {\em Proc. SPIE}
  {\bf 7013},  701328 (July 2008).

\bibitem{2008SPIE.7013E..3AW}
{Wilson}, J.~C., {Hinz}, P.~M., {Skrutskie}, M.~F., {Jones}, T., {Solheid}, E.,
  {Leisenring}, J., {Garnavich}, P., {Kenworthy}, M., {Nelson}, M.~J., and
  {Woodward}, C.~E., ``{LMIRcam: an L/M-band imager for the LBT combined
  focus},'' in [{\em Optical and Infrared
  Interferometry}{\nolinebreak\hspace{0.1em}]},  {\em Proc. SPIE} {\bf 7013},
  70133A (July 2008).

\bibitem{2010SPIE.7735E..3HS}
{Skrutskie}, M.~F., {Jones}, T., {Hinz}, P., {Garnavich}, P., {Wilson}, J.,
  {Nelson}, M., {Solheid}, E., {Durney}, O., {Hoffmann}, W., {Vaitheeswaran},
  V., {McMahon}, T., {Leisenring}, J., and {Wong}, A., ``{The Large Binocular
  Telescope mid-infrared camera (LMIRcam): final design and status},'' in [{\em
  Ground-based and Airborne Instrumentation for Astronomy
  III}{\nolinebreak\hspace{0.1em}]},  {\em Proc. SPIE} {\bf 7735},  77353H
  (July 2010).

\bibitem{2012SPIE.8446E..4FL}
{Leisenring}, J.~M., {Skrutskie}, M.~F., {Hinz}, P.~M., {Skemer}, A., {Bailey},
  V., {Eisner}, J., {Garnavich}, P., {Hoffmann}, W.~F., {Jones}, T.,
  {Kenworthy}, M., {Kuzmenko}, P., {Meyer}, M., {Nelson}, M., {Rodigas}, T.~J.,
  {Wilson}, J.~C., and {Vaitheeswaran}, V., ``{On-sky operations and
  performance of LMIRcam at the Large Binocular Telescope},'' in [{\em
  Ground-based and Airborne Instrumentation for Astronomy
  IV}{\nolinebreak\hspace{0.1em}]},  {\em Proc. SPIE} {\bf 8446},  84464F
  (Sept. 2012).

\bibitem{2014SPIE.9147E..1OH}
{Hoffmann}, W.~F., {Hinz}, P.~M., {Defr{\`e}re}, D., {Leisenring}, J.~M.,
  {Skemer}, A.~J., {Arbo}, P.~A., {Montoya}, M., and {Mennesson}, B.,
  ``{Operation and performance of the mid-infrared camera, NOMIC, on the Large
  Binocular Telescope},'' in [{\em Ground-based and Airborne Instrumentation
  for Astronomy V}{\nolinebreak\hspace{0.1em}]},  {\em Proc. SPIE} {\bf 9147},
  91471O (July 2014).

\bibitem{2015AJ....149..175C}
{Conrad}, A., {de Kleer}, K., {Leisenring}, J., {La Camera}, A., {Arcidiacono},
  C., {Bertero}, M., {Boccacci}, P., {Defr{\`e}re}, D., {de Pater}, I., {Hinz},
  P., {Hofmann}, K.-H., {K{\"u}rster}, M., {Rathbun}, J., {Schertl}, D.,
  {Skemer}, A., {Skrutskie}, M., {Spencer}, J., {Veillet}, C., {Weigelt}, G.,
  and {Woodward}, C.~E., ``{Spatially Resolved M-band Emission from Io's Loki
  Patera-Fizeau Imaging at the 22.8 m LBT},'' {\em AJ}~{\bf 149},  175 (May
  2015).

\bibitem{2016ApJ...817..166S}
{Skemer}, A.~J., {Morley}, C.~V., {Zimmerman}, N.~T., {Skrutskie}, M.~F.,
  {Leisenring}, J., {Buenzli}, E., {Bonnefoy}, M., {Bailey}, V., {Hinz}, P.,
  {Defr{\'e}re}, D., {Esposito}, S., {Apai}, D., {Biller}, B., {Brandner}, W.,
  {Close}, L., {Crepp}, J.~R., {De Rosa}, R.~J., {Desidera}, S., {Eisner}, J.,
  {Fortney}, J., {Freedman}, R., {Henning}, T., {Hofmann}, K.-H., {Kopytova},
  T., {Lupu}, R., {Maire}, A.-L., {Males}, J.~R., {Marley}, M., {Morzinski},
  K., {Oza}, A., {Patience}, J., {Rajan}, A., {Rieke}, G., {Schertl}, D.,
  {Schlieder}, J., {Stone}, J., {Su}, K., {Vaz}, A., {Visscher}, C.,
  {Ward-Duong}, K., {Weigelt}, G., and {Woodward}, C.~E., ``{The LEECH
  Exoplanet Imaging Survey: Characterization of the Coldest Directly Imaged
  Exoplanet, GJ 504 b, and Evidence for Superstellar Metallicity},'' {\em
  ApJ}~{\bf 817},  166 (Feb. 2016).

\bibitem{2004SPIE.5382..742G}
{Gassler}, W., {Herbst}, T.~M., {Ragazzoni}, R., {Andersen}, D.~R.,
  {Arcidiacono}, C., {Baumeister}, H., {Beckmann}, U., {Bertram}, T.,
  {Bizenberger}, P., {Bohnhardt}, H., {Diolaiti}, E., {Eckart}, A., {Farinato},
  J., {Ligori}, S., {Rix}, H.-W., {Rohloff}, R.-R., {Salinari}, P., {Soci}, R.,
  {Straubmeier}, C., {Vernet-Viard}, E., {Weigelt}, G., {Weiss}, R., and {Xu},
  W., ``{LINC-NIRVANA: first attempt of an instrument for a 23-m-class
  telescope},'' in [{\em Second Backaskog Workshop on Extremely Large
  Telescopes}{\nolinebreak\hspace{0.1em}]},  {Ardeberg}, A.~L. and {Andersen},
  T., eds., {\em Proc. SPIE} {\bf 5382},  742--747 (July 2004).

\bibitem{2014SPIE.9147E..1MH}
{Herbst}, T.~M., {Ragazzoni}, R., {Eckart}, A., and {Weigelt}, G., ``{The
  LINC-NIRVANA high resolution imager: challenges from the lab to first
  light},'' in [{\em Ground-based and Airborne Instrumentation for Astronomy
  V}{\nolinebreak\hspace{0.1em}]},  {\em Proc. SPIE} {\bf 9147},  91471M (July
  2014).

\bibitem{2014AAS...22334820C}
{Crepp}, J.~R., {Bechter}, A., {Bechter}, E., {Berg}, M., {Carroll}, J.,
  {Collins}, K., {Corpuz}, T., {Ketterer}, R., {Kielb}, E., {Stoddard}, R.,
  {Eisner}, J.~A., {Gaudi}, B.~S., {Hinz}, P., {Kratter}, K.~M., {Macela}, G.,
  {Quirrenbach}, A., {Skrutskie}, M.~F., {Sozzetti}, A., {Woodward}, C.~E., and
  {Zhao}, B., ``{iLocater: A Diffraction-Limited Doppler Spectrometer for the
  Large Binocular Telescope},'' in [{\em American Astronomical Society Meeting
  Abstracts \#223}{\nolinebreak\hspace{0.1em}]},  {\em American Astronomical
  Society Meeting Abstracts} {\bf 223},  348.20 (Jan. 2014).

\bibitem{2014SPIE.9149E..16V}
{Veillet}, C., {Brynnel}, J., {Hill}, J., {Wagner}, R., {Ashby}, D.,
  {Christou}, J., {Little}, J., and {Summers}, D., ``{LBTO's long march to full
  operation - step 1},'' in [{\em Observatory Operations: Strategies,
  Processes, and Systems V}{\nolinebreak\hspace{0.1em}]},  {\em Proc. SPIE}
  {\bf 9149},  914916 (Aug. 2014).

\bibitem{2006SPIE.6267E..10R}
{Ragazzoni}, R., {Giallongo}, E., {Pasian}, F., {Baruffolo}, A., {Bertram}, R.,
  {Diolaiti}, E., {Di Paola}, A., {Farinato}, J., {Gentile}, G., {Hill}, J.,
  {Lombini}, M., {Pedichini}, F., {Speziali}, R., {Smareglia}, R., and
  {Vernet}, E., ``{The wide-field eyes of the Large Binocular Telescope},'' in
  [{\em Society of Photo-Optical Instrumentation Engineers (SPIE) Conference
  Series}{\nolinebreak\hspace{0.1em}]},  {\em Proc. SPIE} {\bf 6267},  626710
  (June 2006).

\bibitem{2008SPIE.7014E..4TS}
{Speziali}, R., {Di Paola}, A., {Giallongo}, E., {Pedichini}, F., {Ragazzoni},
  R., {Testa}, V., {Baruffolo}, A., {De Santis}, C., {Diolaiti}, E.,
  {Farinato}, J., {Fontana}, A., {Gallozzi}, S., {Gasparo}, F., {Gentile}, G.,
  {Grazian}, A., {Manzato}, P., {Pasian}, F., {Smareglia}, R., and {Vernet},
  E., ``{The Large Binocular Camera: description and performances of the first
  binocular imager},'' in [{\em Ground-based and Airborne Instrumentation for
  Astronomy II}{\nolinebreak\hspace{0.1em}]},  {\em Proc. SPIE} {\bf 7014},
  70144T (July 2008).

\bibitem{2008A&A...482..349G}
{Giallongo}, E., {Ragazzoni}, R., {Grazian}, A., {Baruffolo}, A., {Beccari},
  G., {de Santis}, C., {Diolaiti}, E., {di Paola}, A., {Farinato}, J.,
  {Fontana}, A., {Gallozzi}, S., {Gasparo}, F., {Gentile}, G., {Green}, R.,
  {Hill}, J., {Kuhn}, O., {Pasian}, F., {Pedichini}, F., {Radovich}, M.,
  {Salinari}, P., {Smareglia}, R., {Speziali}, R., {Testa}, V., {Thompson}, D.,
  {Vernet}, E., and {Wagner}, R.~M., ``{The performance of the blue prime focus
  large binocular camera at the large binocular telescope},'' {\em A\&A}~{\bf
  482},  349--357 (Apr. 2008).

\bibitem{1972MNRAS.160P..13W}
{Wynne}, C.~G., ``{Improved three-lens field correctors for paraboloids},''
  {\em MNRAS}~{\bf 160},  13P (1972).

\bibitem{2016AAS...22724012R}
{Rochais}, T.~B., {Rothberg}, B., and {Kuhn}, O., ``{Are the Youngsters Home? A
  Search for Young Clusters in the Merger Remnant NGC 2655},'' in [{\em
  American Astronomical Society Meeting
  Abstracts}{\nolinebreak\hspace{0.1em}]},  {\em American Astronomical Society
  Meeting Abstracts} {\bf 227},  240.12 (Jan. 2016).

\bibitem{2008SPIE.7012E..1MH}
{Hill}, J.~M., {Ragazzoni}, R., {Baruffolo}, A., {Biddick}, C.~J., {Kuhn},
  O.~P., {Diolaiti}, E., {Thompson}, D., and {Rakich}, A., ``{Prime focus
  active optics with the Large Binocular Telescope},'' in [{\em Ground-based
  and Airborne Telescopes II}{\nolinebreak\hspace{0.1em}]},  {\em Proc. SPIE}
  {\bf 7012},  70121M (July 2008).

\bibitem{1999rto..book.....W}
{Wilson}, R.~N.,  [{\em {Reflecting Telescope Optics
  II}}{\nolinebreak\hspace{0.1em}]} (1999).

\bibitem{TechnicalReport...INAF...2014}
{Stangalini}, M., {Pedichini}, F., and {Giallongo}, title =~"{Technical Report:
  LBC Wavefront Reconstructio Software upgrade}", i. . I. m. . a. y. . . p. .~.
  tech. rep.

\bibitem{2014SPIE.9152E..2ES}
{Summers}, K.~R., {Di Paola}, A., {Centrone}, M., {Edwards}, M.~L., {Hill},
  J.~M., {Kuhn}, O.~P., {Pedichini}, F., and {Summers}, D.~M., ``{LBT prime
  focus camera (LBC) control software upgrades},'' in [{\em Software and
  Cyberinfrastructure for Astronomy III}{\nolinebreak\hspace{0.1em}]},  {\em
  Proc. SPIE} {\bf 9152},  91522E (July 2014).

\bibitem{2006SPIE.6269E..0IP}
{Pogge}, R.~W., {Atwood}, B., {Belville}, S.~R., {Brewer}, D.~F., {Byard},
  P.~L., {DePoy}, D.~L., {Derwent}, M.~A., {Eastwood}, J., {Gonzalez}, R.,
  {Krygier}, A., {Marshall}, J.~R., {Martini}, P., {Mason}, J.~A., {O'Brien},
  T.~P., {Osmer}, P.~S., {Pappalardo}, D.~P., {Steinbrecher}, D.~P., {Teiga},
  E.~J., and {Weinberg}, D.~H., ``{The multi-object double spectrographs for
  the Large Binocular Telescope},'' in [{\em Society of Photo-Optical
  Instrumentation Engineers (SPIE) Conference
  Series}{\nolinebreak\hspace{0.1em}]},  {\em Proc. SPIE} {\bf 6269},  62690I
  (June 2006).

\bibitem{2010SPIE.7735E..0AP}
{Pogge}, R.~W., {Atwood}, B., {Brewer}, D.~F., {Byard}, P.~L., {Derwent},
  M.~A., {Gonzalez}, R., {Martini}, P., {Mason}, J.~A., {O'Brien}, T.~P.,
  {Osmer}, P.~S., {Pappalardo}, D.~P., {Steinbrecher}, D.~P., {Teiga}, E.~J.,
  and {Zhelem}, R., ``{The multi-object double spectrographs for the Large
  Binocular Telescope},'' in [{\em Ground-based and Airborne Instrumentation
  for Astronomy III}{\nolinebreak\hspace{0.1em}]},  {\em Proc. SPIE} {\bf
  7735},  77350A (July 2010).

\bibitem{2012SPIE.8446E..0GP}
{Pogge}, R.~W., {Atwood}, B., {O'Brien}, T.~P., {Byard}, P.~L., {Derwent},
  M.~A., {Gonzalez}, R., {Martini}, P., {Mason}, J.~A., {Osmer}, P.~S.,
  {Pappalardo}, D.~P., {Zhelem}, R., {Stoll}, R.~A., {Steinbrecher}, D.~P.,
  {Brewer}, D.~F., {Colarosa}, C., and {Teiga}, E.~J., ``{On-sky performance of
  the Multi-Object Double Spectrograph for the Large Binocular Telescope},'' in
  [{\em Ground-based and Airborne Instrumentation for Astronomy
  IV}{\nolinebreak\hspace{0.1em}]},  {\em Proc. SPIE} {\bf 8446},  84460G
  (Sept. 2012).

\bibitem{2008SPIE.7021E..08A}
{Atwood}, B., {Jorden}, P., and {Pool}, P., ``{The 123 mm 8k{$\times$}3k
  e2v/Ohio State CCD231-68 for MODS},'' in [{\em High Energy, Optical, and
  Infrared Detectors for Astronomy III}{\nolinebreak\hspace{0.1em}]},  {\em
  Proc. SPIE} {\bf 7021},  702108 (July 2008).

\bibitem{2006SPIE.6269E..1JM}
{Marshall}, J.~L., {O'Brien}, T.~P., {Atwood}, B., {Byard}, P.~L., {DePoy},
  D.~L., {Derwent}, M., {Eastman}, J.~D., {Gonzalez}, R., {Pappalardo}, D.~P.,
  and {Pogge}, R.~W., ``{An image motion compensation system for the
  multi-object double spectrograph},'' in [{\em Society of Photo-Optical
  Instrumentation Engineers (SPIE) Conference
  Series}{\nolinebreak\hspace{0.1em}]},  {\em Proc. SPIE} {\bf 6269},  62691J
  (June 2006).

\bibitem{2014SPIE.9151E..4BR}
{Reynolds}, R.~O., {Derwent}, M., {Power}, J., {Kuhn}, O., {Thompson}, D.,
  {O'Brien}, T.~P., {Pogge}, R.~W., and {Wagner}, R.~M., ``{The instrument
  focal plane mask program at the Large Binocular Telescope},'' in [{\em
  Advances in Optical and Mechanical Technologies for Telescopes and
  Instrumentation}{\nolinebreak\hspace{0.1em}]},  {\em Proc. SPIE} {\bf 9151},
  91514B (July 2014).

\bibitem{2015IAUGA..2257946R}
{Rothberg}, B., {Fischer}, J., {Rodrigues}, M., and {Pirzkal}, N., ``{A Monster
  At Any Other Epoch: Are Intermediate Redshift ULIRGs the Progenitors of QSO
  Host Galaxies?},'' {\em IAU General Assembly}~{\bf 22},  2257946 (Aug. 2015).

\bibitem{1988ApJ...325...74S}
{Sanders}, D.~B., {Soifer}, B.~T., {Elias}, J.~H., {Madore}, B.~F., {Matthews},
  K., {Neugebauer}, G., and {Scoville}, N.~Z., ``{Ultraluminous infrared
  galaxies and the origin of quasars},'' {\em ApJ}~{\bf 325},  74--91 (Feb.
  1988).

\bibitem{2013ApJ...767...72R}
{Rothberg}, B., {Fischer}, J., {Rodrigues}, M., and {Sanders}, D.~B.,
  ``{Unveiling the {$\sigma$}-discrepancy. II. Revisiting the Evolution of
  ULIRGs and the Origin of Quasars},'' {\em ApJ}~{\bf 767},  72 (Apr. 2013).

\bibitem{2003SPIE.4841..962S}
{Seifert}, W., {Appenzeller}, I., {Baumeister}, H., {Bizenberger}, P.,
  {Bomans}, D., {Dettmar}, R.-J., {Grimm}, B., {Herbst}, T., {Hofmann}, R.,
  {Juette}, M., {Laun}, W., {Lehmitz}, M., {Lemke}, R., {Lenzen}, R., {Mandel},
  H., {Polsterer}, K., {Rohloff}, R.-R., {Schuetze}, A., {Seltmann}, A.,
  {Thatte}, N.~A., {Weiser}, P., and {Xu}, W., ``{LUCIFER: a Multi-Mode NIR
  Instrument for the LBT},'' in [{\em Instrument Design and Performance for
  Optical/Infrared Ground-based Telescopes}{\nolinebreak\hspace{0.1em}]},
  {Iye}, M. and {Moorwood}, A.~F.~M., eds., {\em Proc. SPIE} {\bf 4841},
  962--973 (Mar. 2003).

\bibitem{2010SPIE.7735E..1LA}
{Ageorges}, N., {Seifert}, W., {J{\"u}tte}, M., {Knierim}, V., {Lehmitz}, M.,
  {Germeroth}, A., {Buschkamp}, P., {Polsterer}, K., {Pasquali}, A., {Naranjo},
  V., {Gemperlein}, H., {Hill}, J., {Feiz}, C., {Hofmann}, R., {Laun}, W.,
  {Lederer}, R., {Lenzen}, R., {Mall}, U., {Mandel}, H., {M{\"u}ller}, P.,
  {Quirrenbach}, A., {Sch{\"a}ffner}, L., {Storz}, C., and {Weiser}, P.,
  ``{LUCIFER1 commissioning at the LBT},'' in [{\em Ground-based and Airborne
  Instrumentation for Astronomy III}{\nolinebreak\hspace{0.1em}]},  {\em Proc.
  SPIE} {\bf 7735},  77351L (July 2010).

\bibitem{2012SPIE.8446E..5LB}
{Buschkamp}, P., {Seifert}, W., {Polsterer}, K., {Hofmann}, R., {Gemperlein},
  H., {Lederer}, R., {Lehmitz}, M., {Naranjo}, V., {Ageorges}, N., {Kurk}, J.,
  {Eisenhauer}, F., {Rabien}, S., {Honsberg}, M., and {Genzel}, R., ``{LUCI in
  the sky: performance and lessons learned in the first two years of
  near-infrared multi-object spectroscopy at the LBT},'' in [{\em Ground-based
  and Airborne Instrumentation for Astronomy IV}{\nolinebreak\hspace{0.1em}]},
  {\em Proc. SPIE} {\bf 8446},  84465L (Sept. 2012).

\bibitem{2010SPIE.7736E..0ER}
{Rabien}, S., {Ageorges}, N., {Barl}, L., {Beckmann}, U., {Bl{\"u}mchen}, T.,
  {Bonaglia}, M., {Borelli}, J.~L., {Brynnel}, J., {Busoni}, L., {Carbonaro},
  L., {Davies}, R., {Deysenroth}, M., {Durney}, O., {Elberich}, M., {Esposito},
  S., {Gasho}, V., {G{\"a}ssler}, W., {Gemperlein}, H., {Genzel}, R., {Green},
  R., {Haug}, M., {Hart}, M.~L., {Hubbard}, P., {Kanneganti}, S., {Masciadri},
  E., {Noenickx}, J., {Orban de Xivry}, G., {Peter}, D., {Quirrenbach}, A.,
  {Rademacher}, M., {Rix}, H.~W., {Salinari}, P., {Schwab}, C., {Storm}, J.,
  {Str{\"u}der}, L., {Thiel}, M., {Weigelt}, G., and {Ziegleder}, J., ``{ARGOS:
  the laser guide star system for the LBT},'' in [{\em Adaptive Optics Systems
  II}{\nolinebreak\hspace{0.1em}]},  {\em Proc. SPIE} {\bf 7736},
  77360E--77360E--12 (July 2010).

\bibitem{2014SPIE.9148E..1BR}
{Rabien}, S., {Barl}, L., {Beckmann}, U., {Bonaglia}, M., {Borelli}, J.~L.,
  {Brynnel}, J., {Buschkamp}, P., {Busoni}, L., {Christou}, J., {Connot}, C.,
  {Davies}, R., {Deysenroth}, M., {Esposito}, S., {G{\"a}ssler}, W.,
  {Gemperlein}, H., {Hart}, M., {Kulas}, M., {Lefebvre}, M., {Lehmitz}, M.,
  {Mazzoni}, T., {Nussbaum}, E., {Orban de Xivry}, G., {Peter}, D.,
  {Quirrenbach}, A., {Raab}, W., {Rahmer}, G., {Storm}, J., and {Ziegleder},
  J., ``{Status of the ARGOS project},'' in [{\em Adaptive Optics Systems
  IV}{\nolinebreak\hspace{0.1em}]},  {\em Proc. SPIE} {\bf 9148},  91481B (July
  2014).

\bibitem{2014SPIE.9149E..2AR}
{Rahmer}, G., {Lefebvre}, M., {Christou}, J., {Raab}, W., {Rabien}, S.,
  {Ziegleder}, J., {Borelli}, J.~L., and {G{\"a}ssler}, W., ``{Early laser
  operations at the Large Binocular Telescope Observatory},'' in [{\em
  Observatory Operations: Strategies, Processes, and Systems
  V}{\nolinebreak\hspace{0.1em}]},  {\em Proc. SPIE} {\bf 9149},  91492A (Aug.
  2014).

\bibitem{2004SPIE.5492.1243H}
{Hofmann}, R., {Gemperlein}, H., {Grimm}, B., {Jutte}, M., {Mandel}, H.,
  {Polsterer}, K., and {Weisz}, H., ``{The cryogenic MOS unit for LUCIFER},''
  in [{\em Ground-based Instrumentation for
  Astronomy}{\nolinebreak\hspace{0.1em}]},  {Moorwood}, A.~F.~M. and {Iye}, M.,
  eds., {\em Proc. SPIE} {\bf 5492},  1243--1253 (Sept. 2004).

\bibitem{2010SPIE.7735E..79B}
{Buschkamp}, P., {Hofmann}, R., {Gemperlein}, H., {Polsterer}, K., {Ageorges},
  N., {Eisenhauer}, F., {Lederer}, R., {Honsberg}, M., {Haug}, M., {Eibl}, J.,
  {Seifert}, W., and {Genzel}, R., ``{The LUCIFER MOS: a full cryogenic mask
  handling unit for a near-infrared multi-object spectrograph},'' in [{\em
  Ground-based and Airborne Instrumentation for Astronomy
  III}{\nolinebreak\hspace{0.1em}]},  {\em Proc. SPIE} {\bf 7735},  773579
  (July 2010).

\bibitem{2014SPIE.9145E..02H}
{Hill}, J.~M., {Ashby}, D.~S., {Brynnel}, J.~G., {Christou}, J.~C., {Little},
  J.~K., {Summers}, D.~M., {Veillet}, C., and {Wagner}, R.~M., ``{The Large
  Binocular Telescope: binocular all the time},'' in [{\em Ground-based and
  Airborne Telescopes V}{\nolinebreak\hspace{0.1em}]},  {\em Proc. SPIE} {\bf
  9145},  914502 (July 2014).

\bibitem{2012SPIE.8444E..1AH}
{Hill}, J.~M., {Green}, R.~F., {Ashby}, D.~S., {Brynnel}, J.~G., {Cushing},
  N.~J., {Little}, J.~K., {Slagle}, J.~H., and {Wagner}, R.~M., ``{The Large
  Binocular Telescope},'' in [{\em Ground-based and Airborne Telescopes
  IV}{\nolinebreak\hspace{0.1em}]},  {\em Proc. SPIE} {\bf 8444},  84441A
  (Sept. 2012).

\end{thebibliography}
\end{document}